

%
%
%
\def\unredoffs{} \def\redoffs{\voffset=-.31truein\hoffset=-.59truein}
\def\speclscape{\special{ps: landscape}}
%
%
%
%
\newbox\leftpage \newdimen\fullhsize \newdimen\hstitle \newdimen\hsbody
\tolerance=1000\hfuzz=2pt
\catcode`\@=11 
\def\bigans{b }
\def\answ{b }

%
\ifx\answ\bigans\message{(This will come out unreduced.}
\magnification=1200\unredoffs\baselineskip=16pt plus 2pt minus 1pt
\hsbody=\hsize \hstitle=\hsize 
\else\message{(This will be reduced.} \let\l@r=L
\magnification=1000\baselineskip=16pt plus 2pt minus 1pt \vsize=7truein
\redoffs \hstitle=8truein\hsbody=4.75truein\fullhsize=10truein\hsize=\hsbody
\output={\ifnum\pageno=0 
  \shipout\vbox{\speclscape{\hsize\fullhsize\makeheadline}
    \hbox to \fullhsize{\hfill\pagebody\hfill}}\advancepageno
  \else
  \almostshipout{\leftline{\vbox{\pagebody\makefootline}}}\advancepageno
  \fi}
\def\almostshipout#1{\if L\l@r \count1=1 \message{[\the\count0.\the\count1]}
      \global\setbox\leftpage=#1 \global\let\l@r=R
 \else \count1=2
  \shipout\vbox{\speclscape{\hsize\fullhsize\makeheadline}
      \hbox to\fullhsize{\box\leftpage\hfil#1}}  \global\let\l@r=L\fi}
\fi
%
\newcount\yearltd\yearltd=\year\advance\yearltd by -1900

\def\Title#1#2{\nopagenumbers\abstractfont\hsize=\hstitle\rightline{#1}%
\vskip 1in\centerline{\titlefont #2}\abstractfont\vskip .5in\pageno=0}
\def\Date#1{\vfill\leftline{#1}\tenpoint\supereject\global\hsize=\hsbody%
\footline={\hss\tenrm\folio\hss}}
%

\def\draftmode{\message{ DRAFTMODE }\def\draftdate{{\rm preliminary draft:
\number\month/\number\day/\number\yearltd\ \ \hourmin}}%
\headline={\hfil\draftdate}\writelabels\baselineskip=20pt plus 2pt minus 2pt
 {\count255=\time\divide\count255 by 60 \xdef\hourmin{\number\count255}
  \multiply\count255 by-60\advance\count255 by\time
  \xdef\hourmin{\hourmin:\ifnum\count255<10 0\fi\the\count255}}}
\def\nolabels{\def\wrlabeL##1{}\def\eqlabeL##1{}\def\reflabeL##1{}}
\def\writelabels{\def\wrlabeL##1{\leavevmode\vadjust{\rlap{\smash%
{\line{{\escapechar=` \hfill\rlap{\sevenrm\hskip.03in\string##1}}}}}}}%
\def\eqlabeL##1{{\escapechar-1\rlap{\sevenrm\hskip.05in\string##1}}}%
\def\reflabeL##1{\noexpand\llap{\noexpand\sevenrm\string\string\string##1}}}
\nolabels
%
\global\newcount\secno \global\secno=0
\global\newcount\meqno \global\meqno=1
\def\newsec#1{\global\advance\secno by1\message{(\the\secno. #1)}
\global\subsecno=0\eqnres@t\noindent{\bf\the\secno. #1}
\writetoca{{\secsym} {#1}}\par\nobreak\medskip\nobreak}
\def\eqnres@t{\xdef\secsym{\the\secno.}\global\meqno=1\bigbreak\bigskip}
\def\sequentialequations{\def\eqnres@t{\bigbreak}}\xdef\secsym{}
\global\newcount\subsecno \global\subsecno=0
\def\subsec#1{\global\advance\subsecno by1\message{(\secsym\the\subsecno. #1)}
\ifnum\lastpenalty>9000\else\bigbreak\fi
\noindent{\it\secsym\the\subsecno. #1}\writetoca{\string\quad
{\secsym\the\subsecno.} {#1}}\par\nobreak\medskip\nobreak}
\def\appendix#1#2{\global\meqno=1\global\subsecno=0\xdef\secsym{\hbox{#1.}}
\bigbreak\bigskip\noindent{\bf Appendix #1. #2}\message{(#1. #2)}
\writetoca{Appendix {#1.} {#2}}\par\nobreak\medskip\nobreak}
%
%
\def\eqnn#1{\xdef #1{(\secsym\the\meqno)}\writedef{#1\leftbracket#1}%
\global\advance\meqno by1\wrlabeL#1}
\def\eqna#1{\xdef #1##1{\hbox{$(\secsym\the\meqno##1)$}}
\writedef{#1\numbersign1\leftbracket#1{\numbersign1}}%
\global\advance\meqno by1\wrlabeL{#1$\{\}$}}
\def\eqn#1#2{\xdef #1{(\secsym\the\meqno)}\writedef{#1\leftbracket#1}%
\global\advance\meqno by1$$#2\eqno#1\eqlabeL#1$$}
%
\newskip\footskip\footskip14pt plus 1pt minus 1pt 
\def\footnotefont{\ninepoint}\def\f@t#1{\footnotefont #1\@foot}
\def\f@@t{\baselineskip\footskip\bgroup\footnotefont\aftergroup\@foot\let\next}
\setbox\strutbox=\hbox{\vrule height9.5pt depth4.5pt width0pt}
\global\newcount\ftno \global\ftno=0
\def\foot{\global\advance\ftno by1\footnote{$^{\the\ftno}$}}
%
\newwrite\ftfile
\def\footend{\def\foot{\global\advance\ftno by1\chardef\wfile=\ftfile
$^{\the\ftno}$\ifnum\ftno=1\immediate\openout\ftfile=foots.tmp\fi%
\immediate\write\ftfile{\noexpand\smallskip%
\noexpand\item{f\the\ftno:\ }\pctsign}\findarg}%
\def\footatend{\vfill\eject\immediate\closeout\ftfile{\parindent=20pt
\centerline{\bf Footnotes}\nobreak\bigskip\input foots.tmp }}}
\def\footatend{}
%
%
\global\newcount\refno \global\refno=1
\newwrite\rfile
\def\ref{[\the\refno]\nref}
\def\nref#1{\xdef#1{[\the\refno]}\writedef{#1\leftbracket#1}%
\ifnum\refno=1\immediate\openout\rfile=refs.tmp\fi
\global\advance\refno by1\chardef\wfile=\rfile\immediate
\write\rfile{\noexpand\item{#1\ }\reflabeL{#1\hskip.31in}\pctsign}\findarg}
\def\findarg#1#{\begingroup\obeylines\newlinechar=`\^^M\pass@rg}
{\obeylines\gdef\pass@rg#1{\writ@line\relax #1^^M\hbox{}^^M}%
\gdef\writ@line#1^^M{\expandafter\toks0\expandafter{\striprel@x #1}%
\edef\next{\the\toks0}\ifx\next\em@rk\let\next=\endgroup\else\ifx\next\empty%
\else\immediate\write\wfile{\the\toks0}\fi\let\next=\writ@line\fi\next\relax}}
\def\striprel@x#1{} \def\em@rk{\hbox{}}
\def\lref{\begingroup\obeylines\lr@f}
\def\lr@f#1#2{\gdef#1{\ref#1{#2}}\endgroup\unskip}

\def\addref#1{\immediate\write\rfile{\noexpand\item{}#1}} 
\def\startrefs#1{\immediate\openout\rfile=refs.tmp\refno=#1}
\def\xref{\expandafter\xr@f}\def\xr@f[#1]{#1}
\def\refs#1{\count255=1[\r@fs #1{\hbox{}}]}
\def\r@fs#1{\ifx\und@fined#1\message{reflabel \string#1 is undefined.}%
\nref#1{need to supply reference \string#1.}\fi%
\vphantom{\hphantom{#1}}\edef\next{#1}\ifx\next\em@rk\def\next{}%
\else\ifx\next#1\ifodd\count255\relax\xref#1\count255=0\fi%
\else#1\count255=1\fi\let\next=\r@fs\fi\next}
%

%
\newwrite\ffile\global\newcount\figno \global\figno=1
\def\fig{fig.~\the\figno\nfig}
\def\nfig#1{\xdef#1{fig.~\the\figno}%
\writedef{#1\leftbracket fig.\noexpand~\the\figno}%
\ifnum\figno=1\immediate\openout\ffile=figs.tmp\fi\chardef\wfile=\ffile%
\immediate\write\ffile{\noexpand\medskip\noexpand\item{Fig.\ \the\figno. }
\reflabeL{#1\hskip.55in}\pctsign}\global\advance\figno by1\findarg}
\def\xfig{\expandafter\xf@g}\def\xf@g fig.\penalty\@M\ {}
\def\figs#1{figs.~\f@gs #1{\hbox{}}}
\def\f@gs#1{\edef\next{#1}\ifx\next\em@rk\def\next{}\else
\ifx\next#1\xfig #1\else#1\fi\let\next=\f@gs\fi\next}
\newwrite\lfile
{\escapechar-1\xdef\pctsign{\string\%}\xdef\leftbracket{\string\{}
\xdef\rightbracket{\string\}}\xdef\numbersign{\string\#}}

\def\writestop{\def\writestoppt{\immediate\write\lfile{\string\pageno%
\the\pageno\string\startrefs\leftbracket\the\refno\rightbracket%
\string\def\string\secsym\leftbracket\secsym\rightbracket%
\string\secno\the\secno\string\meqno\the\meqno}\immediate\closeout\lfile}}
\def\writestoppt{}\def\writedef#1{}
\def\seclab#1{\xdef #1{\the\secno}\writedef{#1\leftbracket#1}\wrlabeL{#1=#1}}
\def\subseclab#1{\xdef #1{\secsym\the\subsecno}%
\writedef{#1\leftbracket#1}\wrlabeL{#1=#1}}
\newwrite\tfile \def\writetoca#1{}
\def\leaderfill{\leaders\hbox to 1em{\hss.\hss}\hfill}
\def\writetoc{\immediate\openout\tfile=toc.tmp
   \def\writetoca##1{{\edef\next{\write\tfile{\noindent ##1
   \string\leaderfill {\noexpand\number\pageno} \par}}\next}}}

%
\catcode`\@=12 
%
\edef\tfontsize{\ifx\answ\bigans scaled\magstep3\else scaled\magstep4\fi}
\font\titlerm=cmr10 \tfontsize \font\titlerms=cmr7 \tfontsize
\font\titlermss=cmr5 \tfontsize \font\titlei=cmmi10 \tfontsize
\font\titleis=cmmi7 \tfontsize \font\titleiss=cmmi5 \tfontsize
\font\titlesy=cmsy10 \tfontsize \font\titlesys=cmsy7 \tfontsize
\font\titlesyss=cmsy5 \tfontsize \font\titleit=cmti10 \tfontsize
\skewchar\titlei='177 \skewchar\titleis='177 \skewchar\titleiss='177
\skewchar\titlesy='60 \skewchar\titlesys='60 \skewchar\titlesyss='60
\def\titlefont{\def\rm{\fam0\titlerm}
\textfont0=\titlerm \scriptfont0=\titlerms \scriptscriptfont0=\titlermss
\textfont1=\titlei \scriptfont1=\titleis \scriptscriptfont1=\titleiss
\textfont2=\titlesy \scriptfont2=\titlesys \scriptscriptfont2=\titlesyss
\textfont\itfam=\titleit \def\it{\fam\itfam\titleit}\rm}
 \ifx\answ\bigans\else scaled\magstep1\fi
\ifx\answ\bigans\def\abstractfont{\tenpoint}\else
\font\abssl=cmsl10 scaled \magstep1
\font\absrm=cmr10 scaled\magstep1 \font\absrms=cmr7 scaled\magstep1
\font\absrmss=cmr5 scaled\magstep1 \font\absi=cmmi10 scaled\magstep1
\font\absis=cmmi7 scaled\magstep1 \font\absiss=cmmi5 scaled\magstep1
\font\abssy=cmsy10 scaled\magstep1 \font\abssys=cmsy7 scaled\magstep1
\font\abssyss=cmsy5 scaled\magstep1 \font\absbf=cmbx10 scaled\magstep1
\skewchar\absi='177 \skewchar\absis='177 \skewchar\absiss='177
\skewchar\abssy='60 \skewchar\abssys='60 \skewchar\abssyss='60
\def\abstractfont{\def\rm{\fam0\absrm}
\textfont0=\absrm \scriptfont0=\absrms \scriptscriptfont0=\absrmss
\textfont1=\absi \scriptfont1=\absis \scriptscriptfont1=\absiss
\textfont2=\abssy \scriptfont2=\abssys \scriptscriptfont2=\abssyss
\textfont\itfam=\bigit \def\it{\fam\itfam\bigit}\def\footnotefont{\tenpoint}%
\textfont\slfam=\abssl \def\sl{\fam\slfam\abssl}%
\textfont\bffam=\absbf \def\bf{\fam\bffam\absbf}\rm}\fi
\def\tenpoint{\def\rm{\fam0\tenrm}
\textfont0=\tenrm \scriptfont0=\sevenrm \scriptscriptfont0=\fiverm
\textfont1=\teni  \scriptfont1=\seveni  \scriptscriptfont1=\fivei
\textfont2=\tensy \scriptfont2=\sevensy \scriptscriptfont2=\fivesy
\textfont\itfam=\tenit \def\it{\fam\itfam\tenit}\def\footnotefont{\ninepoint}%
\textfont\bffam=\tenbf \def\bf{\fam\bffam\tenbf}\def\sl{\fam\slfam\tensl}\rm}
\font\ninerm=cmr9 \font\sixrm=cmr6 \font\ninei=cmmi9 \font\sixi=cmmi6
\font\ninesy=cmsy9 \font\sixsy=cmsy6 \font\ninebf=cmbx9
\font\nineit=cmti9 \font\ninesl=cmsl9 \skewchar\ninei='177
\skewchar\sixi='177 \skewchar\ninesy='60 \skewchar\sixsy='60
\def\ninepoint{\def\rm{\fam0\ninerm}
\textfont0=\ninerm \scriptfont0=\sixrm \scriptscriptfont0=\fiverm
\textfont1=\ninei \scriptfont1=\sixi \scriptscriptfont1=\fivei
\textfont2=\ninesy \scriptfont2=\sixsy \scriptscriptfont2=\fivesy
\textfont\itfam=\ninei \def\it{\fam\itfam\nineit}\def\sl{\fam\slfam\ninesl}%
\textfont\bffam=\ninebf \def\bf{\fam\bffam\ninebf}\rm}
%
%

\hyphenation{anom-aly anom-alies coun-ter-term coun-ter-terms}
\def\inv{^{\raise.15ex\hbox{${\scriptscriptstyle -}$}\kern-.05em 1}}

\def\Dsl{\,\raise.15ex\hbox{/}\mkern-13.5mu D} 
\def\dsl{\raise.15ex\hbox{/}\kern-.57em\partial}

\font\bigit=cmti10 scaled \magstep1
\def\lspace{\ifx\answ\bigans{}\else\qquad\fi}
\def\lbspace{\ifx\answ\bigans{}\else\hskip-.2in\fi} 
\def\boxeqn#1{\vcenter{\vbox{\hrule\hbox{\vrule\kern3pt\vbox{\kern3pt
    \hbox{${\displaystyle #1}$}\kern3pt}\kern3pt\vrule}\hrule}}}
\def\mbox#1#2{\vcenter{\hrule \hbox{\vrule height#2in
        \kern#1in \vrule} \hrule}}  
%

\def\darr#1{\raise1.5ex\hbox{$\leftrightarrow$}\mkern-16.5mu #1}

\def\roughly#1{\raise.3ex\hbox{$#1$\kern-.75em\lower1ex\hbox{$\sim$}}}

\let\includefigures=\iftrue
\let\useblackboard=\iftrue
\newfam\black

\includefigures
\message{If you do not have epsf.tex (to include figures),}
\message{change the option at the top of the tex file.}
\input epsf
\def\figin{\epsfcheck\figin}\def\figins{\epsfcheck\figins}
\def\epsfcheck{\ifx\epsfbox\UnDeFiNeD
\message{(NO epsf.tex, FIGURES WILL BE IGNORED)}
\gdef\figin##1{\vskip2in}\gdef\figins##1{\hskip.5in}
\else\message{(FIGURES WILL BE INCLUDED)}%
\gdef\figin##1{##1}\gdef\figins##1{##1}\fi}
\def\DefWarn#1{}
\def\figinsert{\goodbreak\midinsert}
\def\ifig#1#2#3{\DefWarn#1\xdef#1{fig.~\the\figno}
\writedef{#1\leftbracket fig.\noexpand~\the\figno}%
\figinsert\figin{\centerline{#3}}\medskip\centerline{\vbox{
\baselineskip12pt\advance\hsize by -1truein
\noindent\footnotefont{\bf Fig.~\the\figno:} #2}}
\endinsert\global\advance\figno by1}
\else
\def\ifig#1#2#3{\xdef#1{fig.~\the\figno}
\writedef{#1\leftbracket fig.\noexpand~\the\figno}%
\global\advance\figno by1} \fi

\def\id{{1 \kern-.28em {\rm l}}}

\def\K3{{\bf K3}}
\def\journal#1&#2(#3){\unskip, \sl #1\ \bf #2 \rm(19#3) }
\def\andjournal#1&#2(#3){\sl #1~\bf #2 \rm (19#3) }

\def\eg{{\it e.g.}}

\def\tilde{\widetilde}

\def\frac#1#2{{#1\over#2}}

\def\inbar{\,\vrule height1.5ex width.4pt depth0pt}
\def\IC{\relax\hbox{$\inbar\kern-.3em{\rm C}$}}
\def\IR{\relax{\rm I\kern-.18em R}}
\def\IP{\relax{\rm I\kern-.18em P}}

%
%

%
\catcode`\@=11
\def\slash#1{\mathord{\mathpalette\c@ncel{#1}}}
\overfullrule=0pt

\def\LL{{\cal L}}

\def\OO{{\cal O}}

\def\QQ{{\cal Q}}

\def\underrel#1\over#2{\mathrel{\mathop{\kern\z@#1}\limits_{#2}}}

\catcode`\@=12


%

\def\det{{\rm det}}

\def\det{{\rm det}}


\def\p{{\partial}}

\def\ra{{\rightarrow}}

\def\LL{{\cal L}}
\def\tnu{{\tilde \nu}}

\def\ts{{\tilde s}}

\lref\KulaxiziGY{
  M.~Kulaxizi, A.~Parnachev and K.~Schalm,
  ``On Holographic Entanglement Entropy of Charged Matter,''
[arXiv:1208.2937 [hep-th]].
}

\lref\SonEM{
  D.~T.~Son and A.~O.~Starinets,
  ``Hydrodynamics of r-charged black holes,''
JHEP {\bf 0603}, 052 (2006).
[hep-th/0601157].
}

\lref\KulaxiziGY{
  M.~Kulaxizi, A.~Parnachev and K.~Schalm,
  ``On Holographic Entanglement Entropy of Charged Matter,''
JHEP {\bf 1210}, 098 (2012).
[arXiv:1208.2937 [hep-th]].
}

\lref\RyuBV{
  S.~Ryu and T.~Takayanagi,
  ``Holographic derivation of entanglement entropy from AdS/CFT,''
Phys.\ Rev.\ Lett.\  {\bf 96}, 181602 (2006).
[hep-th/0603001].
}

\lref\RyuEF{
  S.~Ryu and T.~Takayanagi,
  ``Aspects of Holographic Entanglement Entropy,''
JHEP {\bf 0608}, 045 (2006).
[hep-th/0605073].
}

\lref\GoykhmanVY{
  M.~Goykhman, A.~Parnachev and J.~Zaanen,
  ``Fluctuations in finite density holographic quantum liquids,''
JHEP {\bf 1210}, 045 (2012).
[arXiv:1204.6232 [hep-th]].
}

\lref\CveticXP{
  M.~Cvetic, M.~J.~Duff, P.~Hoxha, J.~T.~Liu, H.~Lu, J.~X.~Lu, R.~Martinez-Acosta and C.~N.~Pope {\it et al.},
  ``Embedding AdS black holes in ten-dimensions and eleven-dimensions,''
Nucl.\ Phys.\ B {\bf 558}, 96 (1999).
[hep-th/9903214].
}

\lref\GubserQT{
  S.~S.~Gubser and F.~D.~Rocha,
  ``Peculiar properties of a charged dilatonic black hole in $AdS_5$,''
Phys.\ Rev.\ D {\bf 81}, 046001 (2010).
[arXiv:0911.2898 [hep-th]].
}

\lref\DeWolfeAA{
  O.~DeWolfe, S.~S.~Gubser and C.~Rosen,
  ``Fermi Surfaces in Maximal Gauged Supergravity,''
Phys.\ Rev.\ Lett.\  {\bf 108}, 251601 (2012).
[arXiv:1112.3036 [hep-th]].
}

\lref\GubserYB{
  S.~S.~Gubser and J.~Ren,
  ``Analytic fermionic Green's functions from holography,''
[arXiv:1204.6315 [hep-th]].
}

\lref\DeWolfeUV{
  O.~DeWolfe, S.~S.~Gubser and C.~Rosen,
  ``Fermi surfaces in N=4 Super-Yang-Mills theory,''
[arXiv:1207.3352 [hep-th]].
}

\lref\MyersIJ{
  R.~C.~Myers, M.~F.~Paulos and A.~Sinha,
  ``Holographic Hydrodynamics with a Chemical Potential,''
JHEP {\bf 0906}, 006 (2009). [arXiv:0903.2834 [hep-th]].
}

\lref\BuchelDI{
  A.~Buchel, J.~T.~Liu and A.~O.~Starinets,
  ``Coupling constant dependence of the shear viscosity in N=4 supersymmetric Yang-Mills theory,''
Nucl.\ Phys.\ B {\bf 707}, 56 (2005). [hep-th/0406264].
}

\lref\CremoniniEJ{
  S.~Cremonini and P.~Szepietowski,
  ``Generating Temperature Flow for eta/s with Higher Derivatives: From Lifshitz to AdS,''
JHEP {\bf 1202}, 038 (2012). [arXiv:1111.5623 [hep-th]].
}

\lref\PolicastroSE{
  G.~Policastro, D.~T.~Son and A.~O.~Starinets,
  ``From AdS / CFT correspondence to hydrodynamics,''
JHEP {\bf 0209}, 043 (2002). [hep-th/0205052].
}

\lref\SonSD{
  D.~T.~Son and A.~O.~Starinets,
  ``Minkowski space correlators in AdS / CFT correspondence: Recipe and applications,''
JHEP {\bf 0209}, 042 (2002). [hep-th/0205051].
}

\lref\IqbalBY{
  N.~Iqbal and H.~Liu,
  ``Universality of the hydrodynamic limit in AdS/CFT and the membrane paradigm,''
Phys.\ Rev.\ D {\bf 79}, 025023 (2009). [arXiv:0809.3808 [hep-th]].
}

\lref\KovtunEV{
  P.~K.~Kovtun and A.~O.~Starinets,
  ``Quasinormal modes and holography,''
Phys.\ Rev.\ D {\bf 72}, 086009 (2005).
[hep-th/0506184].
}

\lref\KaminskiDH{
  M.~Kaminski, K.~Landsteiner, J.~Mas, J.~P.~Shock and J.~Tarrio,
  ``Holographic Operator Mixing and Quasinormal Modes on the Brane,''
JHEP {\bf 1002}, 021 (2010).
[arXiv:0911.3610 [hep-th]].
}

\lref\LandauLifshitz{
     L.~D.~Landau and E.~M.~Lifshitz,
     ``Course of theoretical physics volume 6: Fluid mechanics,''
     Butterworth-Heinemann, 1987, 2nd ed.
}

\lref\PolicastroTN{
  G.~Policastro, D.~T.~Son and A.~O.~Starinets,
  ``From AdS / CFT correspondence to hydrodynamics. 2. Sound waves,''
JHEP {\bf 0212}, 054 (2002).
[hep-th/0210220].
}

\lref\ExperimentalZS{
  W.~R.~Abel, A.~C.~Anderson and J.~C.~Wheatley,
  ``Propagation of Zero Sound in Liquid He-3 at Low Temperatures,''
  Phys.~Rev.~Lett.~, {\bf 17} 74 (1966)
}

\lref\BaymPethick{
   G.~Baym and C.~Pethick,
   ``Landau Fermi-Liquid Theory: Concepts and Applications,''
   John Wiley and Sons, Inc., New York, 1991
}

\lref\PinesNozieres{
   D.~Pines and P.~Nozieres,
   ``The Theory of Quantum Liquids''
   W.~A.~Benjamin, New York, 1966
}

\lref\LFLreview{
   A.~A.~Abrikosov and I.~M.~Khalatnikov,
   ``The theory of a fermi liquid (the properties of $^3$He at low temperatures),''
   Rep.~Prog.~Phys.~, {\bf 22}, 329 (1959)
}

\lref\DavisonBXA{
  R.~A.~Davison and A.~Parnachev,
  ``Hydrodynamics of cold holographic matter,''
[arXiv:1303.6334 [hep-th]].
}

\lref\KlebanovTB{
  I.~R.~Klebanov and E.~Witten,
Nucl.\ Phys.\ B {\bf 556}, 89 (1999).
[hep-th/9905104].
}

\lref\RozaliRX{
  M.~Rozali, H.~-H.~Shieh, M.~Van Raamsdonk and J.~Wu,
  ``Cold Nuclear Matter In Holographic QCD,''
JHEP {\bf 0801}, 053 (2008).
[arXiv:0708.1322 [hep-th]].
}

\lref\OgawaBZ{
  N.~Ogawa, T.~Takayanagi and T.~Ugajin,
  ``Holographic Fermi Surfaces and Entanglement Entropy,''
JHEP {\bf 1201}, 125 (2012).
[arXiv:1111.1023 [hep-th]].
}

\lref\HuijseEF{
  L.~Huijse, S.~Sachdev and B.~Swingle,
  ``Hidden Fermi surfaces in compressible states of gauge-gravity duality,''
Phys.\ Rev.\ B {\bf 85}, 035121 (2012).
[arXiv:1112.0573 [cond-mat.str-el]].
}

\lref\AnantuaNJ{
  R.~J.~Anantua, S.~A.~Hartnoll, V.~L.~Martin and D.~M.~Ramirez,
  ``The Pauli exclusion principle at strong coupling: Holographic matter and momentum space,''
JHEP {\bf 1303}, 104 (2013).
[arXiv:1210.1590 [hep-th]].
}

\lref\KarchFA{
  A.~Karch, D.~T.~Son and A.~O.~Starinets,
  "Zero Sound from Holography,''
[arXiv:0806.3796 [hep-th]].
}

\lref\KulaxiziKV{
  M.~Kulaxizi and A.~Parnachev,
  ``Comments on Fermi Liquid from Holography,''
Phys.\ Rev.\ D {\bf 78}, 086004 (2008).
[arXiv:0808.3953 [hep-th]].
}

\lref\KovtunDE{
  P.~Kovtun, D.~T.~Son and A.~O.~Starinets,
  ``Viscosity in strongly interacting quantum field theories from black hole physics,''
Phys.\ Rev.\ Lett.\  {\bf 94}, 111601 (2005).
[hep-th/0405231].
}

\lref\BuchelTZ{
  A.~Buchel and J.~T.~Liu,
  ``Universality of the shear viscosity in supergravity,''
Phys.\ Rev.\ Lett.\  {\bf 93}, 090602 (2004).
[hep-th/0311175].
}

\lref\KulaxiziJX{
  M.~Kulaxizi and A.~Parnachev,
  ``Holographic Responses of Fermion Matter,''
Nucl.\ Phys.\ B {\bf 815}, 125 (2009).
[arXiv:0811.2262 [hep-th]].
}

\lref\HungQK{
  L.~-Y.~Hung and A.~Sinha,
  ``Holographic quantum liquids in 1+1 dimensions,''
JHEP {\bf 1001}, 114 (2010).
[arXiv:0909.3526 [hep-th]].
}

\lref\HoyosBadajozKD{
  C.~Hoyos-Badajoz, A.~O'Bannon and J.~M.~S.~Wu,
  ``Zero Sound in Strange Metallic Holography,''
JHEP {\bf 1009}, 086 (2010).
[arXiv:1007.0590 [hep-th]].
}

\lref\NickelPR{
  D.~Nickel and D.~T.~Son,
  ``Deconstructing holographic liquids,''
New J.\ Phys.\  {\bf 13}, 075010 (2011).
[arXiv:1009.3094 [hep-th]].
}

\lref\LeeEZ{
  B.~-H.~Lee, D.~-W.~Pang and C.~Park,
  ``Zero Sound in Effective Holographic Theories,''
JHEP {\bf 1011}, 120 (2010).
[arXiv:1009.3966 [hep-th]].
}

\lref\BergmanRF{
  O.~Bergman, N.~Jokela, G.~Lifschytz and M.~Lippert,
  ``Striped instability of a holographic Fermi-like liquid,''
JHEP {\bf 1110}, 034 (2011).
[arXiv:1106.3883 [hep-th]].
}

\lref\AmmonHZ{
  M.~Ammon, J.~Erdmenger, S.~Lin, S.~Muller, A.~O'Bannon, J.~P.~Shock, J.~Erdmenger and S.~Lin {\it et al.},
  ``On Stability and Transport of Cold Holographic Matter,''
JHEP {\bf 1109}, 030 (2011).
[arXiv:1108.1798 [hep-th]].
}

\lref\DavisonEK{
  R.~A.~Davison and A.~O.~Starinets,
  ``Holographic zero sound at finite temperature,''
Phys.\ Rev.\ D {\bf 85}, 026004 (2012).
[arXiv:1109.6343 [hep-th]].
}

\lref\DavisonUK{
  R.~A.~Davison and N.~K.~Kaplis,
  ``Bosonic excitations of the $AdS_4$ Reissner-Nordstrom black hole,''
JHEP {\bf 1112}, 037 (2011).
[arXiv:1111.0660 [hep-th]].
}

\lref\GorskyGI{
  A.~Gorsky and A.~V.~Zayakin,
  ``Anomalous Zero Sound,''
JHEP {\bf 1302}, 124 (2013).
[arXiv:1206.4725 [hep-th]].
}

\lref\BrattanNB{
  D.~K.~Brattan, R.~A.~Davison, S.~A.~Gentle and A.~O'Bannon,
  ``Collective Excitations of Holographic Quantum Liquids in a Magnetic Field,''
JHEP {\bf 1211}, 084 (2012).
[arXiv:1209.0009 [hep-th]].
}

\lref\JokelaSE{
  N.~Jokela, M.~Jarvinen and M.~Lippert,
  ``Fluctuations and instabilities of a holographic metal,''
JHEP {\bf 1302}, 007 (2013).
[arXiv:1211.1381 [hep-th]].
}

\lref\PangYPA{
  D.~-W.~Pang,
  ``Probing holographic semi-local quantum liquids with D-branes,''
[arXiv:1306.3816 [hep-th]].
}

\lref\DeyVJA{
  P.~Dey and S.~Roy,
  ``Zero sound in strange metals with hyperscaling violation from holography,''
[arXiv:1307.0195 [hep-th]].
}

\lref\EdalatiTMA{
  M.~Edalati and J.~F.~Pedraza,
  ``Aspects of Current Correlators in Holographic Theories with Hyperscaling Violation,''
[arXiv:1307.0808 [hep-th]].
}

\lref\LiuDM{
  H.~Liu, J.~McGreevy and D.~Vegh,
  ``Non-Fermi liquids from holography,''
Phys.\ Rev.\ D {\bf 83}, 065029 (2011).
[arXiv:0903.2477 [hep-th]].
}

\lref\CubrovicYE{
  M.~Cubrovic, J.~Zaanen and K.~Schalm,
  ``String Theory, Quantum Phase Transitions and the Emergent Fermi-Liquid,''
Science {\bf 325}, 439 (2009).
[arXiv:0904.1993 [hep-th]].
}

\lref\HartnollGU{
  S.~A.~Hartnoll and A.~Tavanfar,
  ``Electron stars for holographic metallic criticality,''
Phys.\ Rev.\ D {\bf 83}, 046003 (2011).
[arXiv:1008.2828 [hep-th]].
}

\lref\SachdevZE{
  S.~Sachdev,
  ``A model of a Fermi liquid using gauge-gravity duality,''
Phys.\ Rev.\ D {\bf 84}, 066009 (2011).
[arXiv:1107.5321 [hep-th]].
}

\lref\FaulknerGT{
  T.~Faulkner and N.~Iqbal,
  ``Friedel oscillations and horizon charge in 1D holographic liquids,''
[arXiv:1207.4208 [hep-th]].
}

\lref\LeeFY{
  S.~-S.~Lee,
  ``TASI Lectures on Emergence of Supersymmetry, Gauge Theory and String in Condensed Matter Systems,''
[arXiv:1009.5127 [hep-th]].
}

\lref\IqbalBF{
  N.~Iqbal and H.~Liu,
  ``Luttinger's Theorem, Superfluid Vortices, and Holography,''
Class.\ Quant.\ Grav.\  {\bf 29}, 194004 (2012).
[arXiv:1112.3671 [hep-th]].
}

\lref\BriganteNU{
  M.~Brigante, H.~Liu, R.~C.~Myers, S.~Shenker and S.~Yaida,
  ``Viscosity Bound Violation in Higher Derivative Gravity,''
Phys.\ Rev.\ D {\bf 77}, 126006 (2008). [arXiv:0712.0805 [hep-th]].
}

\lref\KatsMQ{
  Y.~Kats and P.~Petrov,
  ``Effect of curvature squared corrections in AdS on the viscosity of the dual gauge theory,''
JHEP {\bf 0901}, 044 (2009). [arXiv:0712.0743 [hep-th]].
}

\lref\deHaroXN{
  S.~de Haro, S.~N.~Solodukhin and K.~Skenderis,
  ``Holographic reconstruction of space-time and renormalization in the AdS / CFT correspondence,''
Commun.\ Math.\ Phys.\  {\bf 217}, 595 (2001). [hep-th/0002230].
}

\lref\AstefaneseiWZ{
  D.~Astefanesei, N.~Banerjee and S.~Dutta,
  ``(Un)attractor black holes in higher derivative AdS gravity,''
JHEP {\bf 0811}, 070 (2008). [arXiv:0806.1334 [hep-th]].
}

\lref\CremoniniIH{
  S.~Cremonini, J.~T.~Liu and P.~Szepietowski,
  ``Higher Derivative Corrections to R-charged Black Holes: Boundary Counterterms and the Mass-Charge Relation,''
JHEP {\bf 1003}, 042 (2010). [arXiv:0910.5159 [hep-th]].
}

\lref\GuoEQ{
  Z.~-K.~Guo, N.~Ohta and T.~Torii,
  ``Black Holes in the Dilatonic Einstein-Gauss-Bonnet Theory in Various Dimensions II. Asymptotically AdS Topological Black Holes,''
Prog.\ Theor.\ Phys.\  {\bf 121}, 253 (2009). [arXiv:0811.3068
[gr-qc]].
}

\lref\OhtaLSA{
  N.~Ohta and T.~Torii,
  ``Asymptotically AdS Charged Black Holes in String Theory with Gauss-Bonnet Correction in Various Dimensions,''
[arXiv:1307.3077 [hep-th]].
}

\lref\IqbalIN{
  N.~Iqbal, H.~Liu and M.~Mezei,
  ``Semi-local quantum liquids,''
JHEP {\bf 1204}, 086 (2012).
[arXiv:1105.4621 [hep-th]].
}

\lref\HartnollWM{
  S.~A.~Hartnoll and E.~Shaghoulian,
  ``Spectral weight in holographic scaling geometries,''
JHEP {\bf 1207}, 078 (2012).
[arXiv:1203.4236 [hep-th]].
}

\lref\EdalatiEH{
  M.~Edalati, K.~W.~Lo and P.~W.~Phillips,
  ``Pomeranchuk Instability in a non-Fermi Liquid from Holography,''
Phys.\ Rev.\ D {\bf 86}, 086003 (2012).
[arXiv:1203.3205 [hep-th]].
}

\lref\BalasubramanianBS{
  V.~Balasubramanian, J.~de Boer, V.~Jejjala and J.~Simon,
  ``Entropy of near-extremal black holes in AdS(5),''
JHEP {\bf 0805}, 067 (2008).
[arXiv:0707.3601 [hep-th]].
}

\lref\FareghbalAR{
  R.~Fareghbal, C.~N.~Gowdigere, A.~E.~Mosaffa and M.~M.~Sheikh-Jabbari,
  ``Nearing Extremal Intersecting Giants and New Decoupled Sectors in N = 4 SYM,''
JHEP {\bf 0808}, 070 (2008).
[arXiv:0801.4457 [hep-th]].
}

\lref\JohnstoneEG{
  M.~Johnstone, M.~M.~Sheikh-Jabbari, J.~Simon and H.~Yavartanoo,
  ``Near-Extremal Vanishing Horizon AdS5 Black Holes and Their CFT Duals,''
JHEP {\bf 1304}, 045 (2013).
[arXiv:1301.3387].
}

\lref\JohnstoneIOA{
  M.~Johnstone, M.~M.~Sheikh-Jabbari, J.~Simon and H.~Yavartanoo,
  ``Extremal Black Holes and First Law of Thermodynamics,''
Phys.\ Rev.\ D {\bf 88}, 101503 (2013).
[arXiv:1305.3157 [hep-th]].
}

\lref\NarayanHK{
  K.~Narayan,
  ``On Lifshitz scaling and hyperscaling violation in string theory,''
Phys.\ Rev.\ D {\bf 85}, 106006 (2012).
[arXiv:1202.5935 [hep-th]].
}

\lref\NarayanKS{
  K.~Narayan, T.~Takayanagi and S.~P.~Trivedi,
  ``AdS plane waves and entanglement entropy,''
JHEP {\bf 1304}, 051 (2013).
[arXiv:1212.4328 [hep-th]].
}

\lref\NarayanQGA{
  K.~Narayan,
  ``Non-conformal brane plane waves and entanglement entropy,''
Phys.\ Lett.\ B {\bf 726}, 370 (2013).
[arXiv:1304.6697 [hep-th]].
}

\lref\SachdevTJ{
  S.~Sachdev,
  ``Compressible quantum phases from conformal field theories in 2+1 dimensions,''
Phys.\ Rev.\ D {\bf 86}, 126003 (2012).
[arXiv:1209.1637 [hep-th]].
}

\lref\MetlitskiPD{
  M.~A.~Metlitski and S.~Sachdev,
  ``Quantum phase transitions of metals in two spatial dimensions: I. Ising-nematic order,''
Phys.\ Rev.\ B {\bf 82}, 075127 (2010).
[arXiv:1001.1153 [cond-mat.str-el]].
}

\lref\MahajanJZA{
  R.~Mahajan, D.~M.~Ramirez, S.~Kachru and S.~Raghu,
  ``Quantum critical metals in $d=3+1$,''
Phys.\ Rev.\ B {\bf 88}, 115116 (2013).
[arXiv:1303.1587 [cond-mat.str-el]].
}

\lref\FitzpatrickMJA{
  A.~L.~Fitzpatrick, S.~Kachru, J.~Kaplan and S.~Raghu,
  ``Non-Fermi liquid fixed point in a Wilsonian theory of quantum critical metals,''
Phys.\ Rev.\ B {\bf 88}, 125116 (2013).
[arXiv:1307.0004 [cond-mat.str-el]].
}

\lref\FitzpatrickRFA{
  A.~L.~Fitzpatrick, S.~Kachru, J.~Kaplan and S.~Raghu,
  ``Non-Fermi liquid behavior of large $N_B$ quantum critical metals,''
[arXiv:1312.3321 [cond-mat.str-el]].
}

\lref\FitzpatrickXWA{
  A.~L.~Fitzpatrick, S.~Kachru, J.~Kaplan, S.~A.~Kivelson and S.~Raghu,
  ``Slow Fermions in Quantum Critical Metals,''
[arXiv:1402.5413 [cond-mat.str-el]].
}

\lref\DongTDA{
  X.~Dong, S.~McCandlish, E.~Silverstein and G.~Torroba,
  ``Controlled non-Fermi liquids from spacetime dependent couplings,''
[arXiv:1402.5965 [cond-mat.str-el]].
}

 \Title{} {\vbox{\centerline{AdS/CFT and Landau Fermi liquids}
}}
\bigskip

\centerline{\it   Richard A. Davison, Mikhail Goykhman and  Andrei Parnachev}
\bigskip
\smallskip
\centerline{Lorentz Institute for Theoretical Physics, Leiden University}
\centerline{P.O. Box 9506, Leiden 2300RA, The Netherlands}

\vglue .3cm

\bigskip

\let\includefigures=\iftrue
\bigskip
\noindent
We study the field theory dual to a charged gravitational background in which the low temperature entropy scales linearly with the temperature. We exhibit the existence of a sound mode which is described by hydrodynamics, even at energies much larger than the temperature, and explain how this, and other properties of the field theory, are consistent with those of a (3+1)-dimensional Landau Fermi liquid, finely tuned to the Pomeranchuk critical point. We also discuss how one could engineer a higher-derivative gravitational Lagrangian which reproduces the correct low temperature behavior of shear viscosity in a generic Landau Fermi liquid.
\bigskip

\Date{December 2013}

\newsec{Introduction and summary}

\noindent  Fermi liquid theory is one of the very few low energy effective theories at finite density that we understand well.
Since the AdS/CFT correspondence provides a simple description of strongly coupled
field theories at finite density, it is natural to ask whether a Fermi liquid description
can be recovered at low energies in a holographic model.
The possibility of observing a Fermi surface at leading order in the ``$1/N$ expansion"
was first raised in \RozaliRX.
More recently, the logarithmic violation of the entanglement entropy
was suggested as signifying the appearance of a Fermi surface \refs{\OgawaBZ-\HuijseEF}.
In \AnantuaNJ, a certain singular behavior of the current-current
correlator was suggested as an indicator of a Fermi surface.
These models imply that the effective theory at low energy differs from the
conventional Landau Fermi liquid theory, which is characterized (among other things)
by a linear heat capacity at small temperatures and by the validity of Luttinger's theorem.\foot{
Constructing Fermi liquids by the explicit inclusion of charged fermions in
the bulk is another interesting direction
\refs{\LiuDM\CubrovicYE\HartnollGU-\SachdevZE}.}

On the other hand, the discovery of a gapless mode in holographic models
at finite density and zero temperature \KarchFA\ raised the possibility that
the Landau Fermi liquid is not too far off: it contains a similar excitation called zero sound (subsequent work on various aspects of holographic zero
sound includes  \refs{\KulaxiziKV\KulaxiziJX\HungQK\HoyosBadajozKD\NickelPR\LeeEZ\BergmanRF\AmmonHZ\DavisonEK\DavisonUK\GoykhmanVY\GorskyGI\BrattanNB\JokelaSE\PangYPA\DeyVJA-\EdalatiTMA}). It was pointed out in \KulaxiziKV\ that the equality between the speed of holographic zero sound
and the speed of hydrodynamic sound is consistent with the Landau parameters, which control the interaction strength in a Fermi liquid, being parametrically large. In this paper we consider a particular background (reviewed in Section 2) which exhibits a linear heat capacity
at small temperatures and features a holographic zero sound mode.
In Section 3 we show that linearized hydrodynamics is completely sufficient to describe
the holographic zero sound excitation to second order in the derivative expansion: it  should therefore be identified with
the usual hydrodynamic sound mode.
The situation here is similar to  \DavisonBXA, where a different background was considered.

In Section 4 we consider Landau Fermi liquid theory and show that the observations in the holographic model can be explained
by considering a regime of parametrically small Fermi velocity and taking the second Landau parameter
to the stability bound $F_2\ra-5$ corresponding to the Pomeranchuk critical point.
In particular, the quasiparticle lifetime $\tau\sim (F_2+5)$ becomes parametrically small,
and as a result, linearized hydrodynamics is valid for energies that are much larger
than the temperature, as long as they are small compared to the chemical potential. This parametrically small lifetime means that we are pushing the Landau Fermi liquid into a regime where it behaves like a non-Fermi liquid. This choice of parameters also resolves an apparent paradox:
the viscosity/entropy density ratio for any model that involves Einstein-Hilbert
gravity is $\eta/s=1/4\pi$ \refs{\PolicastroSE\KovtunDE\BuchelTZ-\IqbalBY}.
At first sight, the viscosity/entropy ratio of a generic Fermi liquid
diverges like $\eta/s\sim \mu^3/T^3$ for large ratios
of chemical potential $\mu$ and temperature  $T$.
However, as we explain in Section 4, the  coefficient in front of the leading term in $\eta/s$
 vanishes for $F_2=-5$, and so a Fermi liquid description may be compatible
once we tune the Landau parameters accordingly.

This raises the question of whether we can have a holographic
dual of a generic Landau Fermi liquid, where hydrodynamics breaks down at high
energies and collisionless behavior takes over? One clear signature of any such model is that $\eta/s\sim \mu^3/T^3$.
In Section 5, we describe a procedure by which one could construct a model with precisely this behavior, by adding higher
derivative terms to the gravitational Lagrangian.
We discuss our results in Section 6. The appendices contain some technical details related to Sections 3 and 5.

\newsec{The two-charge black hole}

\noindent
Here we briefly review the
dilatonic black hole in $AdS_5$
recently explored in \refs{\GubserQT\DeWolfeAA\GubserYB-\DeWolfeUV}. This black hole is usually referred to as the two-charge black hole,
because 
it arises as a solution to the truncation of IIB supergravity on $AdS_5\times S^5$
where two of the three $U(1)$ charges
are equal and non-vanishing while the
third one is zero \CveticXP.

After this truncation one obtains the action \GubserQT,\DeWolfeUV
%
\eqn\LLgubser{I_0=\frac{1}{16\pi G}\int
d^5x\sqrt{g}\left(R-\frac{1}{2}(\p\phi)^2-\frac{8}{L^2}
e^{\phi/\sqrt{6}}-\frac{4}{L^2}e^{-2\phi/\sqrt{6}}+2e^{2\phi/\sqrt{6}}F_{ab}F^{ab}\right),}
with $\phi$ a scalar field and $F_{ab}$ the field strength of the
Maxwell field $A_{b}$. The 2-charge black hole solution is \GubserQT,\DeWolfeUV
\eqn\bhmetric{ds^2=e^{2a(r)}\left(h(r)dt^2-dx^2-dy^2-dz^2\right)-\frac{e^{2b(r)}}{h(r)}dr^2\,,}
where
\eqn\anb{a(r)=\log\left(\frac{r}{L}\left(1+\frac{Q^2}{r^2}\right)^{\frac{1}{3}}\right)\,,\quad
b(r)=-\log\left(\frac{r}{L}\left(1+\frac{Q^2}{r^2}\right)^{\frac{2}{3}}\right)\,,\quad
h(r)=1-\frac{(r_H^2+Q^2)^2}{(r^2+Q^2)^2}\,,}
\eqn\dilnflux{\phi
(r)=\sqrt{\frac{2}{3}}\log\left(1+\frac{Q^2}{r^2}\right)\,,\quad
A_t(r)=\frac{Q}{2L}\left(1-\frac{r_H^2+Q^2}{r^2+Q^2}\right).}
%


The temperature $T$, chemical potential $\mu$, entropy density $s$,
charge density $\sigma$, energy density $\varepsilon$ and pressure
$P$ of the dual field theory are given in terms of the parameters of
the 2-charge black hole \bhmetric\ via
\eqn\twocharge{\eqalign{T&={r_H\over \pi L^2},\qquad \mu={\sqrt{2}
Q\over L^2},\qquad s= {r_H\over 4 G L^3} (r_H^2+Q^2),\cr
\sigma&={\sqrt{2}Q s\over 2\pi r_H },\qquad
\varepsilon=3P=\frac{3\left(r_H^2+Q^2\right)^2}{16\pi GL^5}.} }
At large densities, this black hole is dual to a semi-local quantum liquid \IqbalIN\ which violates hyperscaling \HartnollWM. See \refs{\BalasubramanianBS\FareghbalAR\JohnstoneEG\JohnstoneIOA\NarayanHK\NarayanKS-\NarayanQGA} for related work.

\newsec{Hydrodynamics}

In common with other field theories, for small amplitude excitations around this equilibrium state with sufficiently small frequencies and momenta, local thermal equilibrium is maintained and one can describe the system using hydrodynamics \LandauLifshitz. In this hydrodynamic limit, the excitations of the system include hydrodynamic sound modes with dispersion relations \PolicastroTN
\eqn\hydrosounddispersionrelationgeneral{\omega=\pm\sqrt{\frac{dP}{d\varepsilon}}k-i\frac{2\eta}{3\left(\varepsilon+P\right)}k^2+\ldots,}
where the ellipsis denotes higher order terms in $k$. The attenuation of this hydrodynamic sound mode is controlled by the viscosity $\eta$ of the field theory. For the two-charge black hole, $\eta=s/4\pi$ \IqbalBY, and thus the hydrodynamic sound mode dispersion relation is
\eqn\twochargeblackholesounddispersion{\eqalign{\omega&=\pm\frac{1}{\sqrt{3}}k-i\frac{r_HL^2}{6\left(r_H^2+Q^2\right)}k^2+\ldots,\cr
&=\pm\frac{1}{\sqrt{3}}k-i\frac{\pi T}{3\left(\mu^2+2\pi^2T^2\right)}k^2+\ldots.}}
Note that the leading $k^2$ contribution to the attenuation vanishes linearly with $T$ in the limit $T\rightarrow0$. This is because $s\sim T$ at low $T$.

We are interested in the two-charge black hole solution in the large density limit $\mu\gg T,\omega,k$. Furthermore, we will assume $\left|\omega\right|\sim k$ from now, since this is true for the sound mode in which we are interested. Within this large density limit, the ratio $k/T$ is still arbitrary. We will now clarify our previous assertion of how ``sufficiently small'' $k$ must be such that hydrodynamics is applicable. In a quasiparticle description, there is a mean free path $l_\eta$ between (thermal) collisions and the regime of applicability of hydrodynamics is then $kl_\eta\ll1$. For the two-charge black hole, we can identify a natural expansion parameter from the dispersion relation \twochargeblackholesounddispersion\
\eqn\taufortwochargeblackhole{l_\eta=\frac{T}{\mu^2+2\pi^2T^2}.}
This suggests that hydrodynamics is valid provided that $k/\mu\ll\left(\mu/T+2\pi^2T/\mu\right)$. Within the large density limit, this inequality is {\it always} satisfied and hence the hydrodynamic result \twochargeblackholesounddispersion\ should be true for arbitrary $k$ within the large density limit. This is intimately related to the fact that $\eta\sim s$ for holographic theories, since from \hydrosounddispersionrelationgeneral\ we can write the range of applicability of hydrodynamics more abstractly as $\left|\omega\right|\sim k\ll l_\eta^{-1}\sim\eta^{-1}\sim s^{-1}$ and $s\sim T$ is always small (in units of $\mu$) in the large density limit. In a Landau Fermi liquid, $l_\eta\sim\mu/T^2$ is the length scale over which individual fermions interact such that a hydrodynamic state is formed. For a generic holographic theory with $\eta\sim s$, the equivalent length scale is always small, indicating that a single-particle description is not applicable.

This applicability of hydrodynamics at arbitrarily low temperatures was shown analytically for the collective excitations of the Reissner-Nordstrom-AdS$_4$ black brane in \DavisonBXA. To verify this for the two-charge black hole, we have calculated
numerically the dispersion relation of the sound mode of the field
theory dual to the two-charge black hole over a large range of
temperatures. The details of this calculation are given in appendix A and the results are shown in figure 1.
\ifig\loc{Numerical results for the imaginary part of the sound dispersion relation. The real part (not shown) is always $\approx k/\sqrt{3}$. {\bf Left}: The sound attenuation at fixed $k/\mu=0.1$ as a function of $T/\mu$. The numerical results (shown as black dots) agree very well with the hydrodynamic dispersion relation \twochargeblackholesounddispersion\ (shown as a red line) down to very small temperatures. {\bf Right}: Logarithmic plot of the sound attenuation at fixed, very small $T/\mu=0.001$ as a function of $k/\mu$. The best fit to the numerical results (shown as black dots) is a straight line of gradient $\approx2.96$ (shown as a black line), indicating a $k^3$ dependence.}
{\epsfxsize2.5in\epsfbox{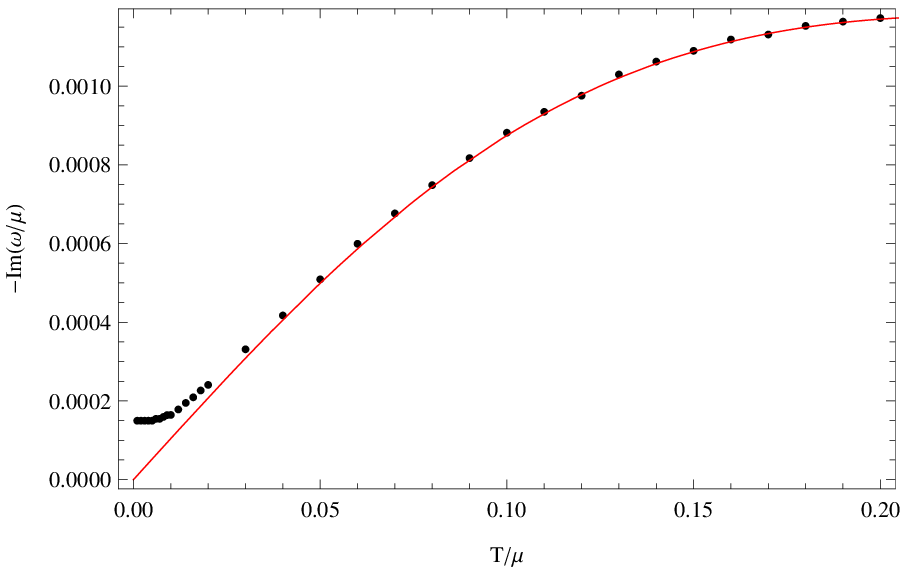} \epsfxsize2.5in\epsfbox{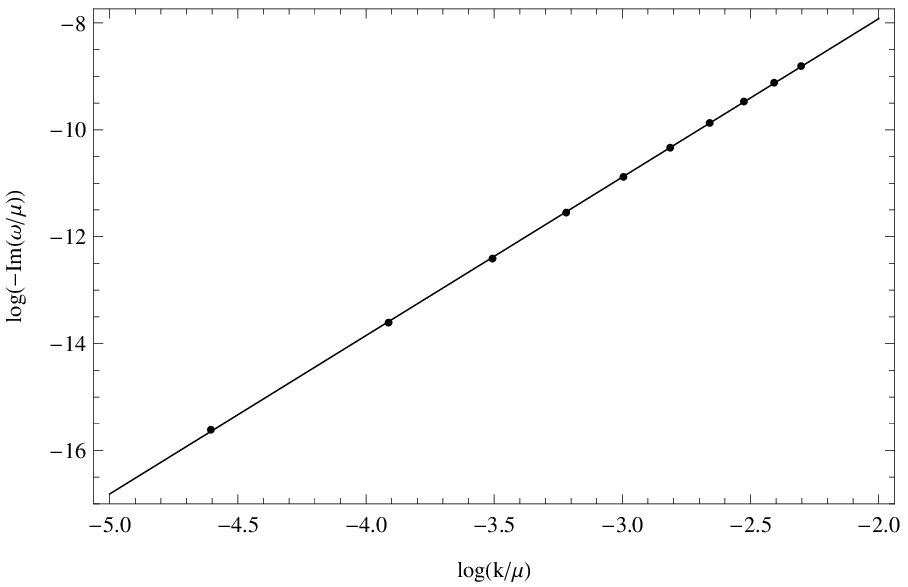}}
The dispersion relation agrees well with the hydrodynamic prediction \twochargeblackholesounddispersion\ down to very low $T$. In the limit $T\rightarrow0$, where the leading order hydrodynamic result \twochargeblackholesounddispersion\ for the sound attenuation vanishes, the numerical result begins to differ from the (vanishing) hydrodynamic prediction. This difference is of order $k^3$ and thus the numerical results are consistent with the hydrodynamic prediction \twochargeblackholesounddispersion\ to $\OO\left(k^2\right)$ for all $T$ in the large density limit. At higher orders in $k$, we expect that they will differ due to logarithmic terms present in the correlators of the semi-local quantum liquid state \DavisonBXA.

We emphasise here that this behaviour is totally different to that in a generic Landau Fermi liquid, which has $\eta\sim T^{-2}$ (since $\eta\ne s/4\pi$). For Landau Fermi liquids, hydrodynamics breaks down at low temperatures $\left|\omega\right|\sim\tau_\eta^{-1}\sim\eta^{-1}\sim T^2/\mu$ where the hydrodynamic sound attenuation becomes very large \refs{\BaymPethick\LFLreview-\PinesNozieres}. It is replaced by a collisionless regime where quantum interactions sustain a `zero sound' mode with a different dispersion relation than \hydrosounddispersionrelationgeneral. This has been confirmed experimentally \ExperimentalZS.

\newsec{Can Landau Fermi liquid theory describe the two-charge black hole?}

\noindent In this section we explore the possibility that Landau Fermi liquid theory might explain the low energy physics of the collective excitations in
the theory holographically dual to \bhmetric. As explained in the previous section, this can only happen for a very specific limit of Fermi liquid theory, since the gravity dual necessarily implies that $\eta/s=1/4\pi$.

\subsec{Establishing the $N$-dependence of the Fermi liquid parameters}

The low temperature heat capacity of the field theory holographically dual to \bhmetric\ is given by
\eqn\cv{  c_V= {\pi L^3 \over 8 G} \mu^2 T = {N^2\over 4}  \mu^2 T,   }
where we have used $L^3/G=2 N^2/\pi$ (see e.g. \SonEM).
This should be compared to the Landau Fermi liquid result
\eqn\cvlfl{  c_V = {k_F m^*\over 3}  T.   }
Let us assume that Luttinger's theorem holds and therefore that the charge density is
\eqn\lutth{   \sigma = \alpha \int_0^{k_F} {d\tau\over (2\pi)^3} = \alpha {k_F^3\over 6\pi^2}.  }
We assume that the numerical coefficient $\alpha$ is $\OO(N^0)$ and counts the number of charged
operators as well as their charge. This would be true, for example, if the charge in the state is carried by an $\OO(1)$ number of fermionic operators. The validity of Luttinger's theorem has been widely debated in the literature (see e.g. \refs{\SachdevZE,\IqbalBF,\FaulknerGT}). In \FaulknerGT\ it was shown that, by including magnetic monopole contributions, Friedel oscillations in the charge density are seen at a wavevector satisfying Luttinger's theorem for a (1+1)-dimensional theory. The validity of this result in higher dimensions (see \SachdevTJ\ for progress in this direction) would be a useful check of our assumption.

With this assumption, we can deduce the $N$-dependence of the various Landau Fermi liquid parameters.
We can rewrite one of the expressions in \twocharge\ as
\eqn\rhoeq{  \sigma = {L^3\over 16\pi G} \mu^3 = {N^2\over8  \pi^2}  \mu^3,  }
at $T=0$. Comparing this to \lutth, we deduce that
\eqn\kf{  k_F = \left({3 N^2\over 4 \alpha}\right)^{1\over3} \mu.  }
Now comparing \cv\ and \cvlfl\ and using \kf\ we obtain
the following expressions for the effective quasiparticle mass and the Fermi velocity
\eqn\effmass{  m^* =    \alpha^{1\over3} \left( {3 N^2  \over4}\right)^{2\over3}   \mu,\qquad \upsilon_F =   \alpha^{-{2\over3}}   \left({ 3 N^2  \over4 }\right)^{-{1\over3}}.    }
We are therefore dealing with an extremely massive Fermi liquid whose Fermi velocity is parametrically small.
We will now compare \effmass\ with the formula for the effective quasiparticle mass in a relativistic Landau Fermi liquid
\eqn\rellflf{  m^* = \mu  \left(  1+{F_1\over 3} \right),  }
to deduce the Landau interaction parameter
\eqn\fone{  {F_1\over 3} =    \alpha^{1\over3} \left( {3 N^2  \over4}\right)^{2\over3} =\alpha^{1\over3} \epsilon^{-2},     }
where we have neglected terms that are $\OO(N^{-2/3})$ and introduced an expansion parameter
\eqn\defx{  \epsilon=  \left( {3 N^2  \over4} \right)^{-{1\over 3}}   \ll 1.   }
We next consider the Landau Fermi liquid expression for the speed of
hydrodynamic sound \eqn\firstsound{  c_1 = {\upsilon_F \over
\sqrt{3} }  \left[ (1+F_0)  \left( 1+ {F_1\over 3}
\right)\right]^{1\over2}  ={1\over\sqrt{3}   },  } where the last
equality follows from the fact that we are dealing with a conformal
field theory where the expectation value of the trace of the
stress-energy tensor vanishes. From \firstsound\ we deduce that
\eqn\fnot{   F_0 = \alpha - 1.} The main lesson of this equation is
that $F_0=\OO(1)$, contrary to $F_1=\OO(N^{4/3})$ as given in
equation \fone.

With these $N$-dependences, we can explain the results of various holographic computations, as we will show shortly. To explain the hydrodynamic results of the previous section, one more assumption is needed. We must tune $F_2\rightarrow-5$ so that the quasiparticles are short-lived. This means that a hydrodynamic, rather than a single-particle, description is applicable at low $T$, as we found for the holographic theory in the previous section.

In this limit, the speed of the zero sound mode $c_0$ is equal to the hydrodynamic value $c_1=1/\sqrt{3}$. To compute the speed of zero sound, one needs to solve the integral equation
\eqn\eqnzsound{   (\frac{c_0}{v_F} - \cos\theta) \nu(\theta,\varphi) = \cos\theta\int F(\theta,\theta')  \nu(\theta',\varphi') {d\Omega'\over2\pi},  }
where $\nu(\theta,\varphi)$ parametrizes the deviation of the Fermi surface from the spherical form.
For zero sound, this deviation is $\varphi$-independent and we can expand in Legendre polynomials,
\eqn\defnul{   \nu(\theta,\varphi) =\sum_l P_l(\cos\theta) \nu_l,\qquad {\tilde \nu}_l={\nu_l\over 2l+1}.  }
Equation \eqnzsound\ then becomes
\eqn\eqnnul{  {\tilde\nu}_l+\sum_l \Omega_{ll'}\left(\frac{c_0}{v_F}\right) F_{l'} {\tilde \nu}_{l'}=0,  }
where
\eqn\defOmega{  \Omega_{ll'} \left(\frac{c_0}{v_F}\right) = {1\over2}  \int_{-1}^1 dy P_l(y) {y\over y-\frac{c_0}{v_F}} P_{l'}(y).  }
We now substitute
\eqn\sdef{ \ts\equiv{\upsilon_F\over c_0}=\OO(\epsilon),   }
 into \eqnnul\ and \defOmega\ and expand in powers of $\ts$. This is an expansion in inverse powers of $N$.
We start by computing the determinant of
\eqn\defA{ A_{ll'} = \delta_{ll'} +\Omega_{ll'}\left(\frac{c_0}{v_F}\right) F_{l'}. }
For \eqnnul\ to have a non-trivial solution, it must be that $\det A=0$.

At this point we will make an important assumption that
\eqn\assumption{  F_n = \OO(1), \qquad n>1,  }
which will simplify things considerably.
We will only keep terms up to $\OO(\ts^2)$  in the matrix $A_{ll'}$ \defA.
After some algebra, we arrive at
\eqn\detmA{ \det A =  1 - {F_0 \ts^2\over 3} - {F_1 \ts^2\over 5} - {F_0 F_1 \ts^2\over9} -  {4\over225} F_1 F_2 \ts^2   +\OO( \ts^2).   }
In \detmA,  we have used \assumption\ and $F_1, F_0 F_1=\OO(\ts^{-2})$ to write only  terms
that can possibly be  $\OO(1)$ (or smaller). It is now clear that keeping terms subleading to $\OO(\ts^2)$ in \defOmega\ and \defA\ will
only modify subleading terms in \detmA.

We now substitute
\eqn\stilde{  \ts^2 =  {3\over (1+F_0) (1+F_1/3)  },    }
into \detmA\ (i.e. impose that $c_0=c_1$) and demand $\det A=0$,  which gives rise to
\eqn\detvanish{   {4\over 25}  { F_1 (5+F_2)\over (1+F_0) (3+F_1)  } =  \OO(\ts^2),   }
which implies
\eqn\ftwo{  F_2=-5+\OO(\ts^2).}

We also need to determine the values of ${\tilde \nu}_l$. This computation is technically more involved than
calculating the determinant: we now need to compute the eigenvector of $A$ associated with the
vanishing eigenvalue. We have done this calculation with $F_n\neq0, n\leq 3$.
The result is
\eqn\nuel{ \tnu_0=1, \quad \tnu_1= {\alpha^{1\over3} \epsilon\over\sqrt{3} },\quad \tnu_2= {2\over 5}, \quad \tnu_3 = \OO(\epsilon^2).   }

\subsec{Observables}
We start by considering observables that present some obvious contradictions with a Landau Fermi liquid description and we will see how these are resolved by counting powers of $N$. We will then  consider the fate of collective excitations at small temperatures, where the Fermi liquid description predicts a particular low temperature behavior.

\bigskip

{\noindent \it Friedel oscillations}

It has been argued that the two-point functions of the currents should develop
a non-analytic behavior at $k=2 k_F$ and $\omega\ra 0$.
As is clear from \kf, such non-analytic structure cannot be observed from the usual linear fluctuations in supergravity, since the equations of motion for these fluctuations have $k\sim\OO(N^0)$. To detect non-analytic behavior at $k_F\sim\OO(N^{2/3})$, one should include the effects of magnetic monopoles along the lines of \FaulknerGT,\SachdevTJ.

\bigskip

{\noindent \it Particle-hole continuum; Landau damping}

A related problem is a non-trivial structure of the spectral function
associated with the particle-hole continuum.
On the $\omega$ axis, the upper edge of the continuum is
defined, for small $\omega$ and $k$, by $\omega = \upsilon_F k$.
The physics of this has been recently reviewed in
\GoykhmanVY, where it was proposed that the continuum is not seen because of a parametrically small $\upsilon_F$.
Precisely such a situation is implied by eq. \effmass. See \AnantuaNJ\ for related work.

\bigskip

{\noindent \it Entanglement entropy}

It has been argued that the presence of a Fermi surface leads to a logarithmic
violation of the area law
\eqn\eelog{    S \simeq   L^2 k_F^2 \log  L +\ldots.    }
Such a violation has not been observed (see e.g. \KulaxiziGY).
The resolution of this puzzle is simple: according to \kf, the $L^2 \log L$ term in
the entanglement entropy must multiply a factor of $N^{4/3}$, while the tree level result
\RyuBV,\RyuEF\ only gives the leading term which is $\OO(N^2)$.

\bigskip

{\noindent \it Collective modes with azimuthal dependence}

As well as zero sound, a Landau Fermi liquid can support other collective modes due to more complicated fluctuations of the Fermi surface. These have not yet been seen in any holographic theory. Including azimuthal dependence in the deformation of the distribution function \defnul\
leads to simple modifications of eqs. \eqnnul\ and \defOmega.
The analog of the matrix \defA, whose determinant must vanish, is now
\eqn\defAm{  A_{ll'}^m = \delta_{ll'} +\Omega_{ll'}^m(s) F_{l'},   }
where
\eqn\defOmegam{  \Omega_{ll'} ^m(s) = {1\over2}  {(l-|m|)! \over (l+|m|)!}  \, \int_{-1}^1 dy P_l^m(y) {y\over y-s} P_{l'}^m(y),  }
and $m\leq l,l'$.
Repeating the steps that led to \detvanish, we obtain
\eqn\detvanishmone{  1 - {3 F_1  (5 + F_2) S_1^2 \over 25 (1 + F_0) (3+F_1)  } = \OO(\ts^2),    }
where $S_1$ is the speed of the $m=1$ collective mode $c^{(m=1)}$ measured in units of $c_0$,  $S_1 = c^{(m=1)}/c_0$.
When $F_0 \sim \alpha \sim \OO(1)$, eq. \detvanishmone\ implies that for generic values of $F_2$, in addition to the $m=0$ zero
sound mode, there is also another collective mode.
However because of the special value of $F_2$ given in \ftwo, there is actually no solution of \detvanishmone\
with $ c^{(m=1)}$ of order one.
We have also checked that for higher modes ($m>1$) the analog of eq. \detvanishmone\ has no solutions
with $ c^{(m>1)}=\OO(1)$.

\bigskip

{\noindent \it Shear viscosity}

In the holographic model that we consider, the shear viscosity
satisfies the universal law \IqbalBY\
\eqn\etas{  {\eta\over s} ={1\over 4\pi}.      }
This, at first sight, presents a puzzle, since the Fermi liquid
result is
\eqn\etafl{   \eta = {1\over5} \rho k_F \upsilon_F \tau_\eta   \sim \rho k_F \upsilon_F \tau,   }
where
\eqn\taueta{ {\tau_\eta\over\tau}   = {2\over\pi^2} \sum_{\nu=1,3,5,\ldots}\;  {2\nu+1\over \nu (\nu+1) [ \nu (\nu+1)-2 \lambda_\eta ] },  }
and $\lambda_\eta=\OO(1)$.
The quasiparticle lifetime, $\tau$ in \etafl\ is related to the scattering probabilities of the quasiparticles $W(\theta,\varphi)$ via
\eqn\taudef{  \tau = {8\pi \over {m^*}^3 \langle W \rangle T^2  },   }
where
\eqn\Wdef{    \langle W \rangle =\int {d\Omega\over 4\pi}  {W(\theta,\varphi)\over \cos(\theta/2)  }.   }
The probabilities are related to the scattering amplitudes which, in turn, can be related to the Landau
parameters, if assumed to be $\varphi$-independent:
\eqn\Wlandau{  W(\theta,\varphi) = 2\pi |A(\theta,\varphi)|^2 = {2 \pi^5\over {m^*}^2 k_F^2} |\sum_l {\cal F}_l P_l(\cos\theta)|^2,   }
where
\eqn\rmFdef{  {\cal F}_l = {F_l\over 1+F_l/(2 l +1)  }.   }
From \rmFdef\ it is clear that generally $ {\cal F}_l =\OO(1)$ and therefore
$\tau \sim k_F^2/(m^* T^2)$, which gives $\eta \sim N^2 \mu^5/T^2$ and
$\eta/s \simeq \mu^3/T^3$ in sharp contradiction with \etas.
However  the value of $F_2$ in \ftwo\ is very non-generic!
Precisely for this value, the denominator in \rmFdef\ vanishes (as $\OO(\epsilon^2)$) and
$\tau$ receives an extra factor of $\epsilon^4$.
Hence, once \ftwo\ is taken into account, the leading order result (in powers of $N^2$) for $\eta/s$
is no longer $\mu^3/T^3$, and one needs to consider higher order terms in $T/\mu$
which might well give rise to the observed value \etas.

\bigskip

{\noindent \it The applicability of hydrodynamics}

In Section 3 we observed that the characteristic length scale where a hydrodynamic description of the sound mode breaks
down is given by eq. \taufortwochargeblackhole. As explained above, this is consistent with Landau Fermi liquid theory
with $F_2=-5$, in which the divergence of the viscosity at low $T$ -- which ensures the breakdown of hydrodynamics -- is suppressed.

\newsec{Higher derivative corrections}

In the previous section we have shown that in the limit
$F_2\rightarrow-5$, many of the generic properties of a Landau Fermi
liquid are no longer realised. It is important that we can recover
these generic properties by altering the gravitational theory such
that $F_2$ is no longer fine-tuned in this way. In this section we
will discuss a procedure by which one may be able to recover the expected $\eta/s\sim\mu^3/T^3$
behavior of a Landau Fermi liquid, by including specific
higher-derivative terms in the gravitational Lagrangian.

\subsec{Entropy}

To compute the entropy of the field theory dual to a gravitational solution, we use the Wald formula
\eqn\Wald{S=-2\pi\int d\Omega\frac{\delta \LL}{\delta
R_{abcd}}E_{ab}E_{cd}\,,}
where $\LL$ is the Lagrangian, $d\Omega=d^3x\sqrt{g_3}$ is the horizon
volume element, $E_{ab}=\sqrt{|g_2|}\epsilon_{ab}$ is the
antisymmetric tensor with indices taking values $t,r$, and
$g_2=g_{tt}g_{rr}$. We will consider a set of three
different four-derivative gravitational terms with coefficients $c_i$, $i=1,2,3$. After setting the
gravitational constant $G=1$ and the AdS radius $L=1$,
the terms in the Lagrangian which
contribute to the Wald formula are
\eqn\genhdact{\LL_R=\frac{1}{16\pi}\left(R+(\beta_1
e^{\gamma_1\phi}+\beta _2 e^{\gamma_2\phi})
(c_1R^2+c_2R_{ab}R^{ab}+c_3R_{abcd}R^{abcd})\right),}
where $\beta_1,\beta_2,\gamma_1$ and $\gamma_2$ are constants. We wish to fix the various constants to satisfy two requirements: that the viscosity is that of a Landau Fermi liquid in the low-temperature limit $r_H/Q\ll 1$, as discussed
in the next subsection, and that the metric and dilaton have the same near-boundary asymptotics as in the two-derivative theory, as discussed in Appendix B.

The Wald formula \Wald\ gives the following expression for the
entropy:
\eqn\genhdentr{S=\frac{A}{4}\left(1+(\beta_1 e^{\gamma_1\phi}+\beta
_2 e^{\gamma_2\phi})\left(2c_1R+c_2(R^t_t+R^r_r)+
4c_3g^{tt}g^{rr}R_{trtr}\right)|_{r=r_H}\right),}
where $A$ is the area of the horizon. For the two-charge black hole \bhmetric\ we obtain
\eqn\hdenrttc{S=\frac{A}{4}\left(1+\frac{8(\beta_1
e^{\gamma_1\phi}+\beta _2
e^{\gamma_2\phi})}{3r_H^2\left(1+\frac{Q^2}{r_H^2}\right)^{2/3}}\left[
(11c_1+4c_2+5c_3)Q^2+24(5c_1+c_2-c_3)r_H^2\right]\right)\,.}
To retain the Fermi liquid result $s\sim T$, we do not want the
entropy to get corrections from these higher-derivative terms. For
this purpose we choose
\eqn\GBc{c_1=1\,,\quad c_2=-4\,,\quad c_3=1\,,}
which corresponds to the Gauss-Bonnet term. The entropy density is
therefore given by \twocharge.

Note that in addition to the explicit corrections to the entropy
density due to the higher-derivative contributions to the Wald
formula, there is also a correction $\Delta s$ coming from the
$\OO(\beta)$ correction to the classical background \bhmetric. The
two-derivative term gives a contribution $s=(g_{xx} (r_H))^{3/2}/4$
to the entropy, and therefore $\Delta s=3\sqrt{g_{xx}(r_H)}\Delta
g_{xx}(r_H)/8$. However, we can define a new radial coordinate
$\rho$ (see Appendix B) such that, as in \MyersIJ,
$g_{xx}(\rho)=-\rho^2$. Then we can choose the constants of
integration of the higher-derivative equations of motion for the
metric such that the horizon radius $\rho_H$ is unchanged, and
therefore $\Delta g_{xx} (\rho_H)=0$ and $\Delta s=0$.

\subsec{Viscosity to entropy ratio}

The total action we study $I=I_0+I_{hd}+I_a+I_{pot}$ is a sum of the
two-derivative terms $I_0$, given by \LLgubser, the
higher-derivative Gauss-Bonnet term with dilaton coupling (see
\genhdact\ and \GBc)
\eqn\lagratwoder{I_{hd}=\frac{1}{16\pi}\int d^5 x \, \sqrt{g}
(\beta_1 e^{\gamma_1\phi}+\beta _2 e^{\gamma_2\phi})
(R^2-4R_{ab}R^{ab}+R_{abcd}R^{abcd})\,,}
the extra dilaton terms
\eqn\Iasympt{I_a=\frac{1}{16\pi}\int d^5x\,\sqrt{g}\left(\beta_1
e^{\gamma_1\phi} (a_1+b_1 g^{\mu\nu}\partial_\mu\phi\partial_\nu\phi)+\beta _2
e^{\gamma_2\phi}(a_2+b_2 g^{\mu\nu}\partial_\mu\phi\partial_\nu\phi)\right)\,,}
and the extra dilaton potential terms
\eqn\extrdilpot{I_{pot}=\frac{1}{16\pi}\int d^5 x\,\sqrt{g}
(g^{\mu\nu}\partial_\mu\phi\partial_\nu\phi)^2
\beta_1e^{\gamma_1\phi}\left(c_1+d_1e^{w_1\phi}+d_2e^{w_2\phi}+d_2e^{w_2\phi}\right)\,.}
Note that the terms \Iasympt\ and \extrdilpot\ do not contribute to the Wald formula \Wald. As we will explain, the terms \Iasympt\ and \extrdilpot\ do not contribute to the viscosity either.

There are two reasons for adding the terms \Iasympt\ and \extrdilpot\ to the action. Firstly, by choosing
\eqn\aobo{a_1=a_2=-120\,,\quad b_1=b_2=4\,,}
the near-boundary asymptotics of the metric and
dilaton are the same as in the two-derivative theory. The additional parameters $c_1$, $d_{1,2,3}$ and $w_{1,2,3}$ are required so that, by fine tuning them, we still have
\eqn\TmurhQ{T\simeq r_H\,,\quad\mu\simeq Q\,,}
in the limit $r_H/Q\ll1$, as in the two-derivative theory \twocharge. We elaborate on both of these points in Appendix B.

Consider the fluctuation $h_x^y(r,t,x)$ of the metric tensor in the momentum representation
\eqn\Ftr{h_x^y=\int \frac{d^2k}{(2\pi)^2}\phi_k(r)e^{-i\omega
t+ikz}\,.}
The action $I=I_0+I_{hd}$, expanded up to second order in
fluctuations, has the general form
\refs{\MyersIJ\BuchelDI-\CremoniniEJ}
\eqn\Squadxy{I{=}\frac{1}{16\pi}\int\frac{ d^2k}{(2\pi)^2}
\left[A(r)\phi_k''\phi_{-k}{+}B(r)\phi
_k'\phi_{-k}'{+}C(r)\phi_k'\phi_{-k}{+}D(r)\phi_k\phi_{-k}{+}E(r)\phi_k''\phi_{-k}''
{+}F(r)\phi_k''\phi_{-k}'\right],}
plus an appropriate Gibbons-Hawking boundary term. The viscosity can be computed from the equation \refs{\MyersIJ\CremoniniEJ}
\eqn\visc{\eta=\frac{1}{8\pi}\left(\kappa_2(r_H)+\kappa_4(r_H)\right),}
where
\eqn\kappas{\kappa_2(r)=\sqrt{-\frac{g_{rr}}{g_{tt}}}\left(A(r)-B(r)+\frac{F'(r)}
{2}\right)\,,\quad
\kappa_4(r)=\left(E(r)\left(\sqrt{-\frac{g_{rr}}{g_{tt}}}\right)'\right)',}
where a prime denotes a derivative with respect to $r$, and where the
momentum $k$ and frequency $\omega$ are set to zero. The terms $I_a$ and $I_{pot}$, given in \Iasympt\ and \extrdilpot, do not involve derivatives of $h^y_x$ and so do not affect the viscosity. We
can separately compute the contribution to the viscosity $\eta_0$
from the two-derivative terms in the action $I_0$, and the
contribution to the viscosity $\eta_1$ from the higher-derivative
terms in the action $I_{hd}$. The total viscosity is given by
$\eta=\eta_0+\eta_1$.

Consider first the higher-derivative term $I_{hd}$. We can expand $I_{hd}$ up to second order in fluctuations and
extract the corresponding coefficients $A_1,B_1,E_1,F_1$, before using equations \visc\ and \kappas\ to give
\eqn\visct{\eqalign{\eta_1&=\frac{1}{8\pi}\left(\kappa_2^{(1)}(r_H)+\kappa_4^{(1)}(r_H)\right)\cr
&=-\frac{\beta_1 (Q^2(1-2\sqrt{6}\gamma_1)+3r_H^2)
}{6\pi}\frac{(r_H^2+Q^2)^{\sqrt{2/3}\gamma_1+1/3}}{r_H^{2\sqrt{2/3}\gamma_1-1/3}}\cr
&-\frac{\beta_2 (Q^2(1-2\sqrt{6}\gamma_2)+3r_H^2)
}{6\pi}\frac{(r_H^2+Q^2)^{\sqrt{2/3}\gamma_2+1/3}}{r_H^{2\sqrt{2/3}\gamma_2-1/3}}.}}

Now we will determine the contribution to the viscosity from the two-derivative
term $I_0$. The most general two-derivative term which contributes
to the entropy density and to the shear viscosity is of the form
\eqn\gentdact{S=\int d^5x \sqrt{g}(R+L_m)\,,}
where $L_m$ is the Lagrangian for matter. We denote the
corresponding viscosity as $\eta_0$ and the corresponding entropy
density as $s_0$. If the classical background is corrected by higher-derivative terms, then the viscosity
receives a correction $\Delta\eta$ and the entropy density receives
a correction $\Delta s$. In addition to these, the higher-derivative
terms, evaluated on the classical background, contribute $\eta_1$
and $s_1$ to the viscosity and the entropy density respectively, as
previously explained.

The question is whether we need to compute $\Delta\eta$ and $\Delta
s$ to determine $\eta /s$. To leading order in $\beta$, the effect of these terms is
\eqn\etasdif{\frac{\eta_0+\Delta\eta+\eta_1}{s_0+\Delta
s+s_1}-\frac{\eta_0+\eta_1}{s_0+s_1}\sim
\frac{\eta_0+\Delta\eta}{s_0+\Delta s}-\frac{\eta_0}{s_0}\sim
s_0\Delta\eta-\eta_0\Delta s.}
But it is known \refs{\PolicastroSE\SonSD-\IqbalBY} that $\eta/s$,
when computed from the two-derivative action \gentdact, does not
depend on the background. In fact since $\eta=g_{xx}^{3/2}/(16\pi)$
(one can derive this in the formalism of \refs{\MyersIJ}, that is
using equation \visc\ for the action \gentdact\ with arbitrary
background metric), and $s=g_{xx}^{3/2}/4$, one always has
$s_0\Delta\eta-\eta_0\Delta s=0$ for any correction to the
background. Therefore as long as we are interested in $\eta/s$, we
do not have to compute $\Delta\eta$ and $\Delta s$
\refs{\MyersIJ,\CremoniniEJ}. In fact, as we noted in the previous subsection, we have $\Delta
s=0$ and $\Delta\eta=0$ satisfied separately, simply because of the
$\Delta g_{xx}(r_H)=0$ equation.

We choose
\eqn\gammabeta{\gamma_1=\frac{7}{2\sqrt{6}}\,,\quad \beta
_1=3\beta\,,\quad\gamma_2=-\frac{7}{\sqrt{6}}\,,\quad
\beta_2=\frac{3}{2}\beta\,.}
The parameters $\beta_{1,2}$ and $\gamma_{1,2}$ then satisfy
\eqn\gbel{\frac{\beta_1}{\beta_2}\frac{\gamma_1}{\gamma _2}=-1\,,}
which, as shown in Appendix B, is crucial to obtain the correct asymptotic behavior of $g_{tt}$ and the dilaton. Therefore in the limit $r_H/Q\ll 1$, we obtain
\eqn\etan{\eta\simeq \frac{Q^5}{r_H^2}.}
This behavior agrees with the low-temperature result of Landau Fermi
liquid theory, provided that the black hole solution has $T\sim r_H$ and $\mu\sim Q$ when $r_H/Q\ll1$, as in the two-derivative theory. This is a non-trivial requirement which needs fine tuning of the coefficients $c_1$, $d_{1,2,3}$ and $w_{1,2,3}$. In Appendix B, we describe a numerical procedure by which one can tune these coefficients and change the leading functional dependence of $T$ and $\mu$ on $r_H$ and $Q$. It is clear from this analysis that it is not easy to produce a gravitational dual of these generic Landau Fermi liquid properties. This further underlines how different holographic metals are, compared to their conventional counterparts.

\newsec{Discussion}

\noindent In Section 4 we described a scenario where a Landau Fermi liquid with
specific parameters was responsible for the properties of the two-charge black hole. The two crucial assumptions were the existence of a single or, at most, $\OO(1)$
Fermi surfaces and the applicability of Luttinger's theorem.
There is some tension with the observation of singularities in the
two-point function of the gauge-invariant fermionic operators \DeWolfeAA,
which would signify that $k_F\sim \mu$ (as opposed to \kf).
If one takes the singularities at $k_F\sim \mu$  seriously, two logical possibilities present themselves:
i) \ the Fermi surface is
formed by the $N^2$ species of gauginos, and Luttinger's theorem is
almost satisfied \DeWolfeAA, or
ii) Luttinger's theorem is strongly violated
in holographic models where charge is sourced by black hole horizons due
to fractionalization in the dual field theory \refs{\HuijseEF,\IqbalBF}.

In this paper we raise a third possibility, which explains all other observable data
(such as the absence of Friedel oscillations, the particle-hole continuum and the logarithmic violation
of entanglement entropy, as well as the behavior of collective excitations).
To make this picture more precise, it would be interesting to identify possible
effects responsible for \kf, and also to check more thoroughly whether one can depart
from the $F_2=-5$ limit and recover a generic Landau Fermi liquid.
In this case, there are precise predictions for the behavior of the sound mode outside
of the hydrodynamic regime.\foot{
We remind the reader that hydrodynamics fixes the speed of the sound mode $c_1$ to
be equal to the thermodynamic value, $c_1 ^2 = \p P/\p \varepsilon$, which is different from the speed
of  zero sound in a generic Fermi liquid.
}
We took the first steps in this direction in Section 5 by describing how one could, in principle, construct a higher derivative gravitational theory
which reproduces the viscosity of a Landau Fermi liquid.

It would be interesting to check our assumption of Luttinger's theorem by a calculation along the lines of \FaulknerGT\ (work in this direction was undertaken in \SachdevTJ). We do not have anything to say at the moment regarding the microscopic origin of the parametrically large $k_F$.
It is known that the naive large $N$ counting breaks down in vector models at finite chemical potential \refs{\LeeFY,\MetlitskiPD}, and the recent works \refs{\MahajanJZA\FitzpatrickMJA\FitzpatrickRFA\FitzpatrickXWA-\DongTDA} have suggested other ways to construct controlled microscopic descriptions of related systems. It would be interesting to see whether the power law in \kf\ can be naturally reproduced.
Finally, let us comment on the nature of the $F_2=-5$ limit. It is known that for $F_l<-2l-1$ the
Landau Fermi liquid becomes unstable. If our scenario is correct, it would be interesting
to identify this instability from a bulk perspective (see \EdalatiEH\ for related work).

\bigskip
\bigskip
{\noindent \bf Acknowledgements:} We would like to thank Mario
Herrero-Valea, Georgios Korpas, Manuela Kulaxizi, Robert Myers, Christopher Rosen, Subir Sachdev and Jan
Zaanen for useful comments and discussions. M.G. thanks the organizers of the
II Postgraduate Meeting On Theoretical Physics, The Institute for
Theoretical Physics (IFT), Madrid, where part of this work was
completed, for hospitality. A.P. thanks IAS, Princeton, where part of
this work was completed, for hospitality. This work was supported in part by a
VIDI grant from NWO.

\appendix{A}{Quasinormal modes of the two-charge black hole}

The dispersion relations of the collective excitations in the field theory dual to a black hole solution are given by the dispersion relations of the quasinormal modes of that black hole \KovtunEV. These quasinormal modes are solutions of the linearised equations of motion for perturbations of the fields around the background solution \bhmetric\ which are infalling at the horizon $r=r_H$ and whose leading term vanishes at the boundary $r\rightarrow\infty$. It is convenient to go to Fourier space and perturb the background fields as follows
\eqn\Fouriertransforms{\eqalign{&\phi\left(r\right)\rightarrow\phi\left(r\right)+\int\frac{d\omega dk}{\left(2\pi\right)^2}e^{-i\omega t+ikz}\varphi\left(r,\omega,k\right),\cr
&A_\mu\left(r\right)\rightarrow A_\mu\left(r\right)+\int\frac{d\omega dk}{\left(2\pi\right)^2}e^{-i\omega t+ikz}a_\mu\left(r,\omega,k\right),\cr
&g_{\mu\nu}\left(r\right)\rightarrow g_{\mu\nu}\left(r\right)+\int\frac{d\omega dk}{\left(2\pi\right)^2}e^{-i\omega t+ikz}h_{\mu\nu}\left(r,\omega,k\right).}}
The equations of motion for these perturbations can be obtained from the action \LLgubser\ but they are very lengthy and we will not present them here.

The sound mode in which we are most interested is associated with longitudinal density oscillations in the field theory and thus is realised in the gravitational dual as a quasinormal mode of $h_{tt}$ (which is dual to the component of the field theory energy-momentum tensor $T^{tt}$). At linear order, $h_{tt}$ couples to $h_{zz}, h_{xx}+h_{yy}, h_{rr}, h_{tr}, h_{zr}, h_{zt}, a_t, a_z, a_r,$ and $\varphi$. Within this set of fields there are, in fact, only three independent degrees freedom which are invariant under infinitesimal diffeomorphisms and U(1) gauge transformations. We choose the basis
\eqn\GIvariables{\eqalign{Z_h&=2\omega kh^z_t+\omega^2h^z_z-k^2hh^t_t-\frac{1}{2}\left[\omega^2-k^2\left(h+\frac{rgh'}{2g+rg'}\right)\right]\left(h^x_x+h^y_y\right),\cr
Z_a&=\omega a_z+ka_t-k\frac{rgA_t'}{2\left(2g+rg'\right)}\left(h^x_x+h^y_y\right),\cr
Z_\varphi&=\varphi-\frac{\sqrt{\frac{3}{2}}rg'}{2\left(2g+rg'\right)}\left(h^x_x+h^y_y\right),}}
and write the equations of motion in the form
\eqn\GIEoMs{\eqalign{&Z_h''+C_1Z_h'+C_2Z_a'+C_3Z_\varphi'+C_4Z_h+C_5Z_a+C_6Z_\varphi=0,\cr
&Z_a''+C_7Z_h'+C_8Z_a'+C_9Z_\varphi'+C_{10}Z_h+C_{11}Z_a+C_{12}Z_\varphi=0,\cr
&Z_\varphi''+C_{13}Z_h'+C_{14}Z_a'+C_{15}Z_\varphi'+C_{16}Z_h+C_{17}Z_a+C_{18}Z_\varphi=0,}}
where the coefficients $C_i$ will be given shortly. Near the spacetime boundary $r\rightarrow0$, the solutions of these equations are of the form
\eqn\nearboundaryexpansions{\eqalign{Z_h&=Z_h^{(1)}\left[\frac{Z_h^{(0)}}{Z_h^{(1)}}\left(1+\ldots\right)+r^{-4}\left(1+\ldots\right)\right],\cr
Z_a&=Z_a^{(1)}\left[\frac{Z_a^{(0)}}{Z_a^{(1)}}\left(1+\ldots\right)+r^{-2}\left(1+\ldots\right)\right],\cr
Z_\varphi&=Z_\varphi^{(1)}\left[\frac{Z_\varphi^{(0)}}{Z_\varphi^{(1)}}r^{-2}\log\left(\frac{r}{r_0}\right)\left(1+\ldots\right)+r^{-2}\left(1+\ldots\right)\right],\cr}}
where the ratios $Z^{(0)}/Z^{(1)}$ are fixed by imposing infalling boundary conditions at the horizon. Quasinormal modes are solutions to these equations of motion which are infalling at the horizon and for which $Z^{(0)}/Z^{(1)}$ vanishes for all three fields. These modes exist only for a discrete set of quasinormal frequencies $\omega\left(k,T,\mu\right)$. To determine these quasinormal frequencies numerically we follow the `determinant method' described in \KaminskiDH\ with one difference. Since the source term in the near-boundary expansion of $\varphi$ is the logarithmic term \KlebanovTB, we extract it by fitting the near-boundary expansion of the numerical solution to the form in \nearboundaryexpansions\ and construct the determinant with this. We note that this method does not give any information about the residue of the pole, only its location in the complex frequency plane. It would be interesting to determine the corresponding residue but we will not address this issue here.

Without further ado, we now list the coefficients $C_i$ in the equations of motion \GIEoMs\
\eqn\GIeqncoefficientsone{\eqalign{C_1=&\Bigl[k^4 \left(r^2-r_H^2\right) \Bigl(8 Q^{10} \left(6 r^4-r^2
   r_H^2+r_H^4\right)+4 Q^8 \left(53 r^6-19 r^4 r_H^2+23 r^2
   r_H^4+3 r_H^6\right)\cr
   &+2 Q^6 \left(262 r^8-181 r^6 r_H^2+83
   r^4 r_H^4+73 r^2 r_H^6+3 r_H^8\right)\cr
   &+Q^4 \left(563
   r^{10}-181 r^8 r_H^2-324 r^6 r_H^4+348 r^4 r_H^6+73 r^2
   r_H^8+r_H^{10}\right)\cr
   &+6 Q^2 r^2 \left(44 r^{10}+13 r^8
   r_H^2-49 r^6 r_H^4+r^4 r_H^6+29 r^2 r_H^8+2
   r_H^{10}\right)\cr
   &+3 r^4 \left(15 r^{10}+15 r^8 r_H^2-16 r^6
   r_H^4-16 r^4 r_H^6+9 r^2 r_H^8+9
   r_H^{10}\right)\Bigr)\cr
   &-2 k^2 \omega ^2 \left(Q^2+r^2\right) \Bigl(2
   Q^{10} \left(5 r^4-5 r^2 r_H^2+2 r_H^4\right)+Q^8 \left(109
   r^6-120 r^4 r_H^2+35 r^2 r_H^4+4 r_H^6\right)\cr
   &+Q^6 \left(303
   r^8-288 r^6 r_H^2+16 r^4 r_H^4+40 r^2
   r_H^6+r_H^8\right)\cr
   &+2 Q^4 r^2 \left(193 r^8-156 r^6 r_H^2-36
   r^4 r_H^4+38 r^2 r_H^6+5 r_H^8\right)\cr
   &+Q^2 r^4 \left(219
   r^8-102 r^6 r_H^2-156 r^4 r_H^4+72 r^2 r_H^6+19
   r_H^8\right)+3 r^6 \left(15 r^8-17 r^4 r_H^4+6
   r_H^8\right)\Bigr)\cr
   &+\omega ^4 \left(Q^2+r^2\right)^4 \left(Q^2+3
   r^2\right)^2 \left(Q^2 \left(6 r^2-2 r_H^2\right)+5
   r^4-r_H^4\right)\Bigr]/\Bigl[r \left(Q^2+r^2\right) \left(Q^2+3 r^2\right)\cr
   &\left(r^2-r_H^2\right) \left(2 Q^2+r^2+r_H^2\right) \left(k^2
   \left(r^2-r_H^2\right) \left(2 Q^2+r^2+r_H^2\right)-\omega ^2
   \left(Q^2+r^2\right)^2\right)\cr
   &\left(k^2 \left(Q^2 \left(4 r^2-2
   r_H^2\right)+3 r^4-r_H^4\right)-\omega ^2 \left(Q^4+4 Q^2 r^2+3
   r^4\right)\right)\Bigr],}}
\eqn\GIeqncoefficientstwo{\eqalign{C_2=-\frac{16 k L Q \left(Q^2+r_H^2\right) \left(k^2 \left(Q^2 \left(2
   r_H^2-4 r^2\right)-3 r^4+r_H^4\right)+\omega ^2 \left(Q^4+4 Q^2
   r^2+3 r^4\right)\right)}{r \left(Q^2+r^2\right) \left(Q^2+3 r^2\right)
   \left(k^2 \left(r^2-r_H^2\right) \left(2
   Q^2+r^2+r_H^2\right)-\omega ^2 \left(Q^2+r^2\right)^2\right)},}}
\eqn\GIeqncoefficientsthree{\eqalign{C_3=0,}}
\eqn\GIeqncoefficientsfour{\eqalign{C_4=&\Bigl[k^4 L^4 \left(Q^2+3 r^2\right) \left(Q^2+r^2\right)^2
   \left(r^2-r_H^2\right) \Bigl(Q^4 \left(8 r^2-4 r_H^2\right)+2 Q^2
   \left(5 r^4+r^2 r_H^2-2 r_H^4\right)\cr
   &+3 r^6+3 r^4 r_H^2-r^2
   r_H^4-r_H^6\Bigr)-2 k^2 \Bigl(Q^{12} \left(r^2 \left(3 L^4
   \omega ^2-32 r_H^2\right)-2 L^4 r_H^2 \omega ^2+32
   r^4\right)\cr
   &+Q^{10} \left(5 r^4 \left(5 L^4 \omega ^2+16 r_H^2\right)-16
   r^2 \left(L^4 r_H^2 \omega ^2+9 r_H^4\right)-L^4 r_H^4
   \omega ^2+64 r^6\right)\cr
   &+Q^8 r^2 \left(r^4 \left(81 L^4 \omega ^2+256
   r_H^2\right)-24 r^2 \left(2 L^4 r_H^2 \omega
   ^2+r_H^4\right)-8 \left(L^4 r_H^4 \omega ^2+32
   r_H^6\right)+24 r^6\right)\cr
   &+4 Q^6 \left(3 r^8 \left(11 L^4 \omega ^2+8
   r_H^2\right)+r^6 \left(96 r_H^4-17 L^4 r_H^2 \omega
   ^2\right)-2 r^4 \left(3 L^4 r_H^4 \omega ^2+32 r_H^6\right)-56
   r^2 r_H^8\right)\cr
   &+Q^4 r^2 \left(115 L^4 r^8 \omega ^2+2 r^6 \left(72
   r_H^4-23 L^4 r_H^2 \omega ^2\right)+r^4 \left(256 r_H^6-34
   L^4 r_H^4 \omega ^2\right)-304 r^2 r_H^8-96
   r_H^{10}\right)\cr
   &+Q^2 r^2 \left(51 L^4 r^{10} \omega ^2-12 L^4 r^8
   r_H^2 \omega ^2+r^6 \left(96 r_H^6-23 L^4 r_H^4 \omega
   ^2\right)+64 r^4 r_H^8-144 r^2 r_H^{10}-16
   r_H^{12}\right)\cr
   &+3 r^4 \left(3 L^4 r^{10} \omega ^2-2 L^4 r^6
   r_H^4 \omega ^2+8 r^4 r_H^8-8 r_H^{12}\right)\Bigr)+L^4
   \omega ^4 \left(Q^2+3 r^2\right)^2
   \left(Q^2+r^2\right)^5\Bigr]/\Bigl[\left(Q^2+r^2\right)^2\cr
   &\left(Q^2+3 r^2\right)
   \left(r^2-r_H^2\right)^2 \left(2 Q^2+r^2+r_H^2\right)^2 \Bigl(k^2
   \left(Q^2 \left(2 r_H^2-4 r^2\right)-3 r^4+r_H^4\right)\cr
   &+\omega ^2
   \left(Q^4+4 Q^2 r^2+3 r^4\right)\Bigr)\Bigr],}}
\eqn\GIeqncoefficientsfive{\eqalign{C_5=&\Bigl[64 k L Q \left(Q^2+r_H^2\right)^3 \Bigl(2 k^4 r^2 \left(2 Q^2+3
   r^2\right) \left(r^2-r_H^2\right) \left(2
   Q^2+r^2+r_H^2\right)\cr
   &+k^2 \omega ^2 \Bigl(Q^6 \left(2 r_H^2-4
   r^2\right)+Q^4 \left(-27 r^4+16 r^2 r_H^2+r_H^4\right)+Q^2
   \left(-40 r^6+18 r^4 r_H^2+8 r^2 r_H^4\right)\cr
   &-15 r^8+9 r^4
   r_H^4\Bigr)+\omega ^4 \left(Q^4+4 Q^2 r^2+3
   r^4\right)^2\Bigr)\Bigr/\Bigl[\left(Q^2+r^2\right)^2 \left(Q^2+3 r^2\right)
   \left(r^2-r_H^2\right)\cr
   &\left(2 Q^2+r^2+r_H^2\right) \left(\omega
   ^2 \left(Q^2+r^2\right)^2-k^2 \left(r^2-r_H^2\right) \left(2
   Q^2+r^2+r_H^2\right)\right) \Bigl(k^2 \Bigl(Q^2 \left(2 r_H^2-4
   r^2\right)\cr
   &-3 r^4+r_H^4\Bigr)+\omega ^2 \left(Q^4+4 Q^2 r^2+3
   r^4\right)\Bigr)\Bigr],}}
\eqn\GIeqncoefficientssix{\eqalign{C_6=-&\Bigl[8 \sqrt{\frac{2}{3}} k^2 Q^2 \left(Q^2+r_H^2\right)^2 \Bigl(k^4
   \left(-\left(r^2-r_H^2\right)^2\right) \left(2
   Q^2+r^2+r_H^2\right)^2 \Bigl(2 Q^2 \left(6 r^2-r_H^2\right)\cr
   &+15
   r^4-r_H^4\Bigr)+2 k^2 \omega ^2 \Bigl(2 Q^8 \left(3 r^4-7 r^2
   r_H^2+2 r_H^4\right)+Q^6 \left(57 r^6-106 r^4 r_H^2+29 r^2
   r_H^4+4 r_H^6\right)\cr
   &+Q^4 \left(110 r^8-158 r^6 r_H^2-13 r^4
   r_H^4+36 r^2 r_H^6+r_H^8\right)\cr
   &+Q^2 r^2 \left(72 r^8-58 r^6
   r_H^2-79 r^4 r_H^4+40 r^2 r_H^6+9 r_H^8\right)+r^4
   \left(15 r^8-29 r^4 r_H^4+10 r_H^8\right)\Bigr)\cr
   &+\omega ^4
   \left(Q^2+r^2\right)^2 \Bigl(2 Q^6 \left(r^2+r_H^2\right)+Q^4 \left(-13
   r^4+32 r^2 r_H^2+r_H^4\right)\cr
   &+Q^2 \left(-42 r^6+54 r^4
   r_H^2+16 r^2 r_H^4\right)-15 r^8+27 r^4
   r_H^4\Bigr)\Bigr)\Bigr]/\Bigl[\left(Q^2+r^2\right)^3 \left(Q^2+3 r^2\right)\cr
   & \left(r^2-r_H^2\right) \left(2 Q^2+r^2+r_H^2\right) \left(\omega
   ^2 \left(Q^2+r^2\right)^2-k^2 \left(r^2-r_H^2\right) \left(2
   Q^2+r^2+r_H^2\right)\right)\cr
   &\left(k^2 \left(Q^2 \left(2 r_H^2-4
   r^2\right)-3 r^4+r_H^4\right)+\omega ^2 \left(Q^4+4 Q^2 r^2+3
   r^4\right)\right)\Bigr],}}
\eqn\GIeqncoefficientsseven{\eqalign{C_7=&-\Bigl[k Q r \left(Q^2+r_H^2\right) \Bigl(k^2 \left(2 Q^4 \left(2
   r^2+r_H^2\right)+Q^2 \left(5 r^4+6 r^2 r_H^2+r_H^4\right)+3
   r^2 \left(r^4+r_H^4\right)\right)\cr
   &+\omega ^2 \left(Q^2-3 r^2\right)
   \left(Q^2+r^2\right)^2\Bigr)\Bigr]/\Bigl[L \left(Q^2+3 r^2\right) \Bigl(\omega ^2
   \left(Q^2+r^2\right)^2\cr
   &-k^2 \left(r^2-r_H^2\right) \left(2
   Q^2+r^2+r_H^2\right)\Bigr) \Bigl(k^2 \left(Q^2 \left(2 r_H^2-4
   r^2\right)-3 r^4+r_H^4\right)\cr
   &+\omega ^2 \left(Q^4+4 Q^2 r^2+3
   r^4\right)\Bigr)\Bigr],}}
\eqn\GIeqncoefficientseight{\eqalign{C_8=&\Bigl[k^2 \left(r^2-r_H^2\right) \Bigl(4 Q^6 \left(r_H^2-5
   r^2\right)+4 Q^4 \left(2 r^4-11 r^2 r_H^2+r_H^4\right)\cr
   &+Q^2
   \left(27 r^6-9 r^4 r_H^2-35 r^2 r_H^4+r_H^6\right)+9 r^2
   \left(r^2-r_H^2\right) \left(r^2+r_H^2\right)^2\Bigr)\cr
   &-\omega ^2
   \left(Q^2+r^2\right)^2 \left(Q^2+3 r^2\right)\left(2 Q^2
   \left(r^2+r_H^2\right)+3 r^4+r_H^4\right)\Bigr]/\Bigl[r \left(Q^2+3
   r^2\right) \left(r^2-r_H^2\right)\cr
   & \left(2 Q^2+r^2+r_H^2\right)
   \left(k^2 \left(r^2-r_H^2\right) \left(2
   Q^2+r^2+r_H^2\right)-\omega ^2 \left(Q^2+r^2\right)^2\right)\Bigr],}}
\eqn\GIeqncoefficientsnine{\eqalign{C_9=\frac{\sqrt{\frac{2}{3}} k Q r \left(Q^2+r_H^2\right)}{L
   \left(Q^2+r^2\right)^2},}}
\eqn\GIeqncoefficientsten{\eqalign{C_{10}=&\Bigl[4 k r^2 \left(2 Q^3+3 Q r^2\right) \left(Q^2+r_H^2\right)^3\Bigr]/\Bigl[L
   \left(Q^2+r^2\right) \left(Q^2+3 r^2\right) \left(r^2-r_H^2\right)\cr
   &\left(2 Q^2+r^2+r_H^2\right) \left(k^2 \left(Q^2 \left(2 r_H^2-4
   r^2\right)-3 r^4+r_H^4\right)+\omega ^2 \left(Q^4+4 Q^2 r^2+3
   r^4\right)\right)\Bigr],}}
\eqn\GIeqncoefficientseleven{\eqalign{C_{11}=\Bigl[&-L^4 \left(Q^4+4 r^2 Q^2+3 r^4\right) \left(r^2-r_H^2\right)^2
   \left(2 Q^2+r^2+r_H^2\right)^2 \left(3 r^4-r_H^4+Q^2 \left(4
   r^2-2 r_H^2\right)\right) k^6\cr
   &-\left(r^2-r_H^2\right) \left(2
   Q^2+r^2+r_H^2\right) \Bigl(2 \left(3 r_H^2 \omega ^2 L^4+64
   r^4-r^2 \left(5 \omega ^2 L^4+64 r_H^2\right)\right) Q^{10}\cr
   &+\left(256
   r^6+\left(64 r_H^2-71 L^4 \omega ^2\right) r^4-40 \left(8
   r_H^4-L^4 r_H^2 \omega ^2\right) r^2+3 L^4 r_H^4 \omega
   ^2\right) Q^8\cr
   &+4 r^2 \left(24 r^6+2 \left(64 r_H^2-23 L^4 \omega
   ^2\right) r^4+\left(23 L^4 r_H^2 \omega ^2-88 r_H^4\right) r^2-64
   r_H^6+5 L^4 r_H^4 \omega ^2\right) Q^6\cr
   &+2 \left(3 \left(32
   r_H^2-37 L^4 \omega ^2\right) r^8+4 \Bigl(11 r_H^2 \omega ^2
   L^4+32 r_H^4\right) r^6+\left(23 L^4 r_H^4 \omega ^2-192
   r_H^6\right) r^4\cr
   &-32 r_H^8 r^2\Bigr) Q^4+\left(-126 L^4 \omega ^2
   r^{10}+6 \left(5 r_H^2 \omega ^2 L^4+16 r_H^4\right) r^8+44 L^4
   r_H^4 \omega ^2 r^6-96 r_H^8 r^4\right) Q^2\cr
   &+3 L^4 r^8 \left(5
   r_H^4-9 r^4\right) \omega ^2\Bigr) k^4+\omega ^2 \Bigl(\left(6
   r_H^2 \omega ^2 L^4+64 r^4+32 r_H^4-8 r^2 \left(\omega ^2 L^4+12
   r_H^2\right)\right) Q^{14}\cr
   &+\left(592 r^6-\left(77 \omega ^2 L^4+832
   r_H^2\right) r^4+8 \left(7 r_H^2 \omega ^2 L^4+18
   r_H^4\right) r^2+96 r_H^6+3 L^4 r_H^4 \omega ^2\right)
   Q^{12}\cr
   &+2 \Bigl(588 r^8-6 \left(25 \omega ^2 L^4+32 r_H^2\right)
   r^6+\left(103 L^4 r_H^2 \omega ^2-928 r_H^4\right) r^4+2
   \left(240 r_H^6+7 L^4 \omega ^2 r_H^4\right) r^2\cr
   &+52
   r_H^8\Bigr) Q^{10}+\Bigl(784 r^{10}+\left(1776 r_H^2-617 L^4
   \omega ^2\right) r^8-128 \left(26 r_H^4-3 L^4 r_H^2 \omega
   ^2\right) r^6\cr
   &+\left(103 L^4 r_H^4 \omega ^2-480 r_H^6\right)
   r^4+1200 r_H^8 r^2+48 r_H^{10}\Bigr) Q^8+2 \Bigl(84 r^{12}+28
   \left(28 r_H^2-13 L^4 \omega ^2\right) r^{10}\cr
   &+\left(193 L^4 r_H^2
   \omega ^2-132 r_H^4\right) r^8-32 \left(49 r_H^6-3 L^4
   r_H^4 \omega ^2\right) r^6+540 r_H^8 r^4+288 r_H^{10} r^2+4
   r_H^{12}\Bigr) Q^6\cr
   &+r^2 \Bigl(96 r_H^{12}+720 r^2
   r_H^{10}-784 r^4 r_H^8+r^{10} \left(336 r_H^2-495 L^4
   \omega ^2\right)+8 r^8 \left(25 r_H^2 \omega ^2 L^4+98
   r_H^4\right)\cr
   &+r^6 \left(193 L^4 r_H^4 \omega ^2-1152
   r_H^6\right)\Bigr) Q^4-2 r^4 \Bigl(-60 r_H^{12}+144 r^4
   r_H^8-50 L^4 r^6 \omega ^2 r_H^4+90 L^4 r^{10} \omega ^2\cr
   &-21 r^8
   \left(r_H^2 \omega ^2 L^4+4 r_H^4\right)\Bigr) Q^2+3 L^4 r^{12}
   \left(7 r_H^4-9 r^4\right) \omega ^2\Bigr) k^2\cr
   &-\left(Q^4+4 r^2 Q^2+3
   r^4\right)^2 \omega ^4 \Bigl(\left(-\omega ^2 L^4+16 r^2-16
   r_H^2\right) Q^8+4 \Bigl(2 r^4+\left(8 r_H^2-L^4 \omega ^2\right)
   r^2\cr
   &-10 r_H^4\Bigr) Q^6+2 \left(-16 r_H^6+8 r^2 r_H^4+r^4
   \left(8 r_H^2-3 L^4 \omega ^2\right)\right) Q^4-4 \left(2 r_H^8-2
   r^4 r_H^4+L^4 r^6 \omega ^2\right) Q^2\cr
   &-L^4 r^8 \omega
   ^2\Bigl)\Bigr]/\Bigl[\left(Q^2+r^2\right) \left(Q^2+3 r^2\right)
   \left(r^2-r_H^2\right)^2 \left(2 Q^2+r^2+r_H^2\right)^2
   \Bigl(\left(Q^2+r^2\right)^2 \omega ^2\cr
   &-k^2 \left(r^2-r_H^2\right)
   \left(2 Q^2+r^2+r_H^2\right)\Bigr) \Bigl(\left(-3 r^4+r_H^4+Q^2
   \left(2 r_H^2-4 r^2\right)\right) k^2\cr
   &+\left(Q^4+4 r^2 Q^2+3 r^4\right)
   \omega ^2\Bigr)\Bigr],}}
\eqn\GIeqncoefficientstwelve{\eqalign{C_{12}=&\Bigl[4 \sqrt{\frac{2}{3}} k Q r^2 \left(Q^2+r_H^2\right) \Bigl(k^4
   \left(-Q^2\right) \left(r^2-r_H^2\right)^2 \left(2
   Q^2+r^2+r_H^2\right)^2 \Bigl(4 Q^4\cr
   &+2 Q^2 \left(3
   r^2+r_H^2\right)+3 r^4+r_H^4\Bigr)-k^2 \omega ^2
   \left(Q^2+r^2\right)^2 \Bigl(2 Q^8 \left(2 r^2+r_H^2\right)\cr
   &+Q^6
   \left(-7 r^4+50 r^2 r_H^2-19 r_H^4\right)+Q^4 \left(-15 r^6+58
   r^4 r_H^2+13 r^2 r_H^4-20 r_H^6\right)\cr
   &+Q^2 \left(-6 r^8+18
   r^6 r_H^2+29 r^4 r_H^4-12 r^2 r_H^6-5 r_H^8\right)+9
   r^6 r_H^4-3 r^2 r_H^8\Bigr)+3 \omega ^4 \left(Q^2+r^2\right)^4\cr
   &\left(Q^6+Q^4 \left(r^2+4 r_H^2\right)+Q^2 \left(-r^4+6 r^2
   r_H^2+2 r_H^4\right)+3 r^2 r_H^4\right)\Bigr)\Bigr]/\Bigl[L
   \left(Q^2+r^2\right)^3\cr
   &\left(Q^2+3 r^2\right) \left(r^2-r_H^2\right)
   \left(2 Q^2+r^2+r_H^2\right) \left(\omega ^2 \left(Q^2+r^2\right)^2-k^2
   \left(r^2-r_H^2\right) \left(2 Q^2+r^2+r_H^2\right)\right)\cr
   &\left(k^2 \left(Q^2 \left(2 r_H^2-4 r^2\right)-3
   r^4+r_H^4\right)+\omega ^2 \left(Q^4+4 Q^2 r^2+3 r^4\right)\right)\Bigr],}}
\eqn\GIeqncoefficientsthirteen{\eqalign{C_{13}&=\Bigl[4 \sqrt{6} Q^2 r \left(k^2-\omega ^2\right)
   \left(Q^2+r^2\right)^3\Big]/\Bigl[\left(Q^2+3 r^2\right) \Bigl(\omega ^2
   \left(Q^2+r^2\right)^2\cr
   &-k^2 \left(r^2-r_H^2\right) \left(2
   Q^2+r^2+r_H^2\right)\Bigr) \Bigl(k^2 \left(Q^2 \left(2 r_H^2-4
   r^2\right)-3 r^4+r_H^4\right)\cr
   &+\omega ^2 \left(Q^4+4 Q^2 r^2+3
   r^4\right)\Bigr)\Bigr],}}
\eqn\GIeqncoefficientsfourteen{\eqalign{C_{14}=\frac{8 \sqrt{6} k L Q \left(Q^2+r^2\right) \left(Q^2+r_H^2\right)}{r
   \left(Q^2+3 r^2\right) \left(k^2 \left(r^2-r_H^2\right) \left(2
   Q^2+r^2+r_H^2\right)-\omega ^2 \left(Q^2+r^2\right)^2\right)},}}
\eqn\GIeqncoefficientsfifteen{\eqalign{C_{15}=\frac{Q^2 \left(6 r^2-2 r_H^2\right)+5 r^4-r_H^4}{r
   \left(r^2-r_H^2\right) \left(2 Q^2+r^2+r_H^2\right)},}}
\eqn\GIeqncoefficientssixteen{\eqalign{C_{16}=-&\Bigl[4 \sqrt{6} Q^2 \left(Q^2+r^2\right)
   \left(Q^2+r_H^2\right)^2\Bigr]/\Bigl[\left(Q^2+3 r^2\right)
   \left(r^2-r_H^2\right) \left(2 Q^2+r^2+r_H^2\right)\cr
   &\left(k^2
   \left(Q^2 \left(2 r_H^2-4 r^2\right)-3 r^4+r_H^4\right)+\omega ^2
   \left(Q^4+4 Q^2 r^2+3 r^4\right)\right)\Bigr],}}
\eqn\GIeqncoefficientsseventeen{\eqalign{C_{17}=&\Bigl[32 \sqrt{6} k L Q^3 \left(k^2-\omega ^2\right) \left(Q^2+r^2\right)
   \left(Q^2+r_H^2\right)\Bigr]/\Bigl[\left(Q^2+3 r^2\right) \Bigl(\omega ^2
   \left(Q^2+r^2\right)^2\cr
   &-k^2 \left(r^2-r_H^2\right) \left(2
   Q^2+r^2+r_H^2\right)\Bigr) \Bigl(k^2 \left(Q^2 \left(2 r_H^2-4
   r^2\right)-3 r^4+r_H^4\right)\cr
   &+\omega ^2 \left(Q^4+4 Q^2 r^2+3
   r^4\right)\Bigr)\Bigr],}}
\eqn\GIeqncoefficientseighteen{\eqalign{C_{18}=&\Bigl[-L^4 \left(Q^4+4 r^2 Q^2+3 r^4\right) \left(r^2-r_H^2\right)^2
   \left(2 Q^2+r^2+r_H^2\right)^2 \left(3 r^4-r_H^4+Q^2 \left(4
   r^2-2 r_H^2\right)\right) k^6\cr
   &+\left(r^2-r_H^2\right) \left(2
   Q^2+r^2+r_H^2\right) \Bigl(2 \left(-3 r_H^2 \omega ^2 L^4+48
   r^4+40 r_H^4+r^2 \left(5 L^4 \omega ^2-88 r_H^2\right)\right)
   Q^{10}\cr
   &+\left(200 r^6+\left(71 L^4 \omega ^2-144 r_H^2\right) r^4-40
   \left(r_H^2 \omega ^2 L^4+5 r_H^4\right) r^2+144 r_H^6-3
   L^4 r_H^4 \omega ^2\right) Q^8\cr
   &+4 \Bigl(73 r^8+\left(46 L^4 \omega ^2-44
   r_H^2\right) r^6-\left(23 r_H^2 \omega ^2 L^4+30
   r_H^4\right) r^4-\left(28 r_H^6+5 L^4 \omega ^2
   r_H^4\right) r^2\cr
   &+29 r_H^8\Bigr) Q^6+\Bigl(324 r^{10}+\left(222
   L^4 \omega ^2-256 r_H^2\right) r^8-8 \left(11 r_H^2 \omega ^2
   L^4+5 r_H^4\right) r^6\cr
   &-2 \left(24 r_H^6+23 L^4 \omega ^2
   r_H^4\right) r^4-28 r_H^8 r^2+48 r_H^{10}\Bigr) Q^4+2
   \Bigl(90 r^{12}+\left(63 L^4 \omega ^2-48 r_H^2\right) r^{10}\cr
   &-\left(15
   r_H^2 \omega ^2 L^4+64 r_H^4\right) r^8+\left(24 r_H^6-22
   L^4 r_H^4 \omega ^2\right) r^6-6 r_H^8 r^4+4
   r_H^{12}\Bigr) Q^2\cr
   &+3 r^6 \left(12 r^8+9 L^4 \omega ^2 r^6-16
   r_H^4 r^4-5 L^4 r_H^4 \omega ^2 r^2+4 r_H^8\right)\Bigr)
   k^4+\left(Q^2+r^2\right) \omega ^2 \Bigl(2 \Bigl(3 r_H^2 \omega ^2
   L^4\cr
   &+8 r^4-16 r_H^4+r^2 \left(8 r_H^2-4 L^4 \omega
   ^2\right)\Bigr) Q^{12}+\Bigl(-216 r^6+\left(720 r_H^2-69 L^4 \omega
   ^2\right) r^4\cr
   &+50 \left(L^4 r_H^2 \omega ^2-12 r_H^4\right) r^2+96
   r_H^6+3 L^4 r_H^4 \omega ^2\Bigr) Q^{10}+\Bigl(-672
   r^8+\left(1216 r_H^2-231 L^4 \omega ^2\right) r^6\cr
   &+4 \left(39
   r_H^2 \omega ^2 L^4+2 r_H^4\right) r^4+\left(25 L^4 r_H^4
   \omega ^2-736 r_H^6\right) r^2+184 r_H^8\Bigr) Q^8+\Bigl(-1096
   r^{10}\cr
   &+2 \left(720 r_H^2-193 L^4 \omega ^2\right) r^8+4 \left(57
   r_H^2 \omega ^2 L^4+64 r_H^4\right) r^6+\left(78 L^4 r_H^4
   \omega ^2-352 r_H^6\right) r^4\cr
   &-344 r_H^8 r^2+96
   r_H^{10}\Bigr) Q^6-2 \Bigl(492 r^{12}+\left(171 L^4 \omega ^2-488
   r_H^2\right) r^{10}-\left(79 r_H^2 \omega ^2 L^4+264
   r_H^4\right) r^8\cr
   &+\left(176 r_H^6-57 L^4 r_H^4 \omega
   ^2\right) r^6+44 r_H^8 r^4+48 r_H^{10} r^2-8
   r_H^{12}\Bigr) Q^4+r^2 \Bigl(-432 r^{12}\cr
   &+3 \left(80 r_H^2-51 L^4
   \omega ^2\right) r^{10}+\left(42 r_H^2 \omega ^2 L^4+488
   r_H^4\right) r^8+\left(79 L^4 r_H^4 \omega ^2-192
   r_H^6\right) r^6-88 r_H^8 r^4\cr
   &-16 r_H^{12}\Bigr) Q^2-3 r^8
   \left(24 r^8+9 L^4 \omega ^2 r^6-40 r_H^4 r^4-7 L^4 r_H^4 \omega
   ^2 r^2+16 r_H^8\right)\Bigr) k^2\cr
   &-\left(Q^2+r^2\right)^4 \omega ^4
   \Bigl(\left(-\omega ^2 L^4+24 r^2-24 r_H^2\right) Q^8-4 \left(3
   r^4+\left(2 L^4 \omega ^2-30 r_H^2\right) r^2+27 r_H^4\right)
   Q^6\cr
   &-2 \left(42 r^6+\left(11 L^4 \omega ^2-60 r_H^2\right) r^4-30
   r_H^4 r^2+48 r_H^6\right) Q^4-12 \Bigl(9 r^8+\left(2 L^4 \omega
   ^2-6 r_H^2\right) r^6\cr
   &-5 r_H^4 r^4+2 r_H^8\Bigr) Q^2-9 r^6
   \left(r^2 \omega ^2 L^4+4 r^4-4
   r_H^4\right)\Bigr)\Bigr]/\Bigl[\left(Q^2+r^2\right) \left(Q^2+3 r^2\right)
   \left(r^2-r_H^2\right)^2\cr
   &\left(2 Q^2+r^2+r_H^2\right)^2
   \left(\left(Q^2+r^2\right)^2 \omega ^2-k^2 \left(r^2-r_H^2\right)
   \left(2 Q^2+r^2+r_H^2\right)\right) \Bigl(\Bigl(-3 r^4+r_H^4\cr
   &+Q^2
   \left(2 r_H^2-4 r^2\right)\Bigr) k^2+\left(Q^4+4 r^2 Q^2+3 r^4\right)
   \omega ^2\Bigr)\Bigr].}}

\appendix{B}{Black hole thermodynamics}

In this appendix we discuss the thermodynamics of the charged black hole solution of the theory whose action is the sum of \LLgubser,
\lagratwoder, \Iasympt\ and \extrdilpot.

\subsec{Adding Gibbons-Hawking boundary terms}

The total action is
\eqn\totact{I=I_0+I_{hd}+I_a+I_{pot},}
and must be supplemented by boundary terms which compensate for the variations of derivatives of the fields on the
boundary of AdS, $r=\infty$. For the two-derivative action $I_0$, this is
the Gibbons-Hawking term
\eqn\GH{I_{GH}=-\frac{1}{8\pi}\int d^4 x\sqrt{|h|}\,K,}
where $h_{ab}$ is the metric on the boundary, $K$ is the trace of the extrinsic curvature tensor $K_a^b=\nabla _an^b$, and the unit vector
normal to the boundary is given by
\eqn\normvec{n^r=\sqrt{g^{rr}}\,,\quad n^a=0\,,\;\; a\neq r.}
For the Gauss-Bonnet term $I_{hd}$, the necessary boundary term is (see
\eg\ \refs{\AstefaneseiWZ, \CremoniniIH})
\eqn\GHGB{I_{bd}=\frac{1}{4\pi}\int d^4x \sqrt{-
h}\,(\beta_1e^{\gamma_1\phi}+\beta
_2e^{\gamma_2\phi})\,\left(KK_a^bK^b_a-\frac{1}{3}(K^3+2K^a_bK^b_cK^c_a)\right).}

Consider first the two-derivative action $I_0$. Evaluating it on the ansatz
\eqn\metrica{ds^2=e^{2a(r)}(h(r)dt^2-dx^2-dy^2-dz^2)-\frac{e^{2b(r)}}{h(r)}dr^2,}
with dilaton $\phi(r)$ and gauge potential $A_t(r)$, we obtain
\eqn\EHanz{\eqalign{I_{0}&=\frac{1}{16\pi}\int d^5 xe^{4 a+ b}
\left(-4 e^{-\sqrt{\frac{2}{3}} \phi }-8 e^{\frac{\phi
}{\sqrt{6}}}+\frac{1}{2} e^{-2 b} h \left(8 a' \left(5 a'-2
b'\right)+\phi ^{\prime 2}+16 a''\right)+\right.\cr &\left.+e^{-2 b}
\left(-4 e^{-2 a+\sqrt{\frac{2}{3}} \phi } A_t^{\prime 2}+\left(9
a'-b'\right) h'+h''\right)\right).}}
Despite the presence of the second derivatives of the fields $h''$
and $a''$ in the action \EHanz, we have a well-defined variational
problem due to the boundary term \GH. Now, evaluating the Gauss-Bonnet term $I_{hd}$ on the ansatz
\metrica, we obtain
\eqn\GBans{\eqalign{I_{hd}&=\frac{1}{16\pi}\int d^5 x 12(\beta_1
e^{\gamma_1\phi}+\beta_2 e^{\gamma_2\phi}) e^{4a-3 b }  a'
\left(a' h^{\prime 2}{+}2 h^2 a' \left(5 a^{\prime 2}{-}4 a' b'{+}4
a''\right){+}\right. \cr &+\left. h \left(h' \left(9 a^{\prime
2}{-}3 a' b'{+}2 a''\right){+}a' h''\right)\right).}}
Again, the action depends on the second derivatives of the fields. Thanks to the boundary terms \GHGB, we have a well-defined
variational problem.

\subsec{Scaling charge}

Following \GubserQT, we note that the total action $I_{tot}$
possesses a scaling symmetry
\eqn\Gscsym{(x,y,z)\rightarrow c(x,y,z)\,,\;\; t\rightarrow
t/c^3\,,\;\; b\rightarrow b+4\log c\,,\;\;a\rightarrow a-\log
c\,,\;\;h\rightarrow c^8h\,,\;\;A_t\rightarrow c^3A_t\,.}
The corresponding conserved charge is given by
\eqn\scch{\QQ=-8 C_0 A_t+ e^{4 a-b} h'+12\beta e^{4 a-3 b} h a' h'
\left(e^{\gamma_1 \phi } \left(a'+2 \gamma_1 \phi '\right)+\chi
e^{\gamma_2 \phi} \left(a'+2 \gamma_2 \phi '\right)\right),}
where we have used the equation of motion for the gauge field
\eqn\Ateq{A_t'= C_0 e^{-2 a+b-\sqrt{2/3} \phi },}
and where $C_0$ is a constant. We have also denoted $\beta_1=\beta$ and
$\beta_2=\chi\beta$. For our choice \gammabeta, $\chi=1/2$. We choose the integration constant, which is related to the field theory charge density, to be
\eqn\CnQ{C_0=Q(Q^2+r_H^2)\,.}
which is consistent with the two-derivative solution.

To $\OO(\beta)$, the charge conservation equation $\QQ(r=r_H)=\QQ(r=\infty)$ can be written
\eqn\ePrel{-8C_0A_t(\infty)+(1+18\beta)r^5h'|_{r=\infty}+48(1+\chi)\beta
(r_H^2+Q^2)^2=16\pi Ts,}
where we have used
\eqn\ePrelu{s=\frac{1}{4}e^{3a(r_H)}\,,\;\;
T=\frac{1}{4\pi}e^{a-b}h'|_{r=r_H}\,,\;\; A_t(r_H)=0\,,}
the asymptotic near-boundary behavior $e^{4a-b}=e^{\delta
}r^5=(1+18\beta)r^5$, and equation \gammabeta; the near-boundary
behavior of $\delta$ is derived in the next subsection. At the
two-derivative level, this is identical to the thermodynamic
relation
\eqn\Pehd{-P=\varepsilon-Ts-\mu\sigma\,,\quad\quad P=\varepsilon/3\,.}
Our strategy will be to evaluate numerically this equation at $\OO(\beta)$ and solve for $A_t\left(\infty\right)$ at this order. This allows us to determine how the chemical potential depends upon $r_H$ and $Q$ at $\OO(\beta)$.

\subsec{Temperature and chemical potential}

In the remainder of this appendix we are going to describe how to numerically construct the black hole solution to $\OO(\beta)$, and then explain how, by fine tuning the parameters $c_1$, $d_{1,2,3}$ and $w_{1,2,3}$ in the extra dilaton potential terms $I_{pot}$, one can change the functional dependence of $T$ and $\mu$ on $r_H$ and $Q$. It is
convenient to choose the ansatz (as in \OhtaLSA)
\eqn\Bdeltaan{ds^2=B(\rho)e^{-2\delta(\rho)}dt^2-\rho^2
(dx^2+dy^2+dz^2)-\frac{d\rho^2}{B(\rho)},}
where $\rho$ is a new radial coordinate. In this ansatz the total
Lagrangian, including the Gibbons-Hawking terms \GH\ and \GHGB, is
\eqn\LtotdB{\eqalign{L_{tot}&{=}\frac{1}{2} e^{-\delta } \rho
\left(2 \rho  \left(-4 e^{-\sqrt{\frac{2}{3}} \phi } \rho -8
e^{\frac{\phi }{\sqrt{6}}} \rho -4 e^{2 \delta +\sqrt{\frac{2}{3}}
\phi} \rho  A_t^{\prime 2}-3 B'\right)+B \left(-12+12 \rho \delta
'+\rho ^2 \phi ^{\prime 2}\right)\right)\cr &+3\beta e^{-\delta
+\gamma_1 \phi } \left(a_1 \rho ^3+B \left(-b_1 \rho ^3 \phi
^{\prime 2}-4 B' \left(1+3 \gamma_1 \rho \phi
'\right)\right.\right.\cr&+\left.\left. B \left(-8 \gamma_1 \phi
'+\left(c_1+d_1 e^{w_1 \phi }+d_2 e^{w_2 \phi }+d_3 e^{w_3 \phi
}\right) \rho ^3 \phi ^{\prime 4}+8 \delta ' \left(1+3 \gamma_1 \rho
\phi '\right)\right)\right)\right)\cr &+ 3 \chi \beta e^{-\delta
+\gamma_2 \phi } \left(a_2 \rho ^3+B \left(-b_2 \rho ^3 \phi
^{\prime 2}-4 B' \left(1+3 \gamma_2 \rho \phi '\right)+B \left(-8
\gamma_2 \phi '+8 \delta ' \left(1+3 \gamma_2 \rho \phi
'\right)\right)\right)\right).}}
It is convenient to work with the dimensionless parameter $\theta=r_H/Q$.

First of all, we solve for the gauge field
\eqn\AtdB{A_t'=C_0 e^{-\delta -\sqrt{\frac{2}{3}} \phi}\rho^{-3},}
with $C_0$ given in \CnQ. Then we solve the equations of motion for $\delta$, $B$ and $\phi$ perturbatively
up to $\OO(\beta)$. Denoting the background (two-derivative) solution
as $\delta_0$, $B_0$ and $\phi_0$, the equations for the $\OO(\beta)$
corrections to the background, which we denote as $\delta_1$, $B_1$
and $\phi _1$, are
\eqn\hoeq{\eqalign{&3 e^{\frac{7 \phi_0}{\sqrt{6}}} \rho\,  B_1
\left(12+\rho ^2 \phi_0^{\prime 2}\right)-\sqrt{6} e^{\frac{7
\phi_0}{\sqrt{6}}} \rho  \phi_1 \left(6 \rho \left(4
e^{\frac{\phi_0}{\sqrt{6}}} \rho -B_0'\right)-B_0 \left(12+\rho ^2
\phi_0^{\prime 2}\right)\right)\cr &+3 \left(3 \left(a_2+2 a_1
e^{\frac{7}{2} \sqrt{\frac{3}{2}} \phi_0}\right) \rho ^3+6
e^{\frac{7 \phi_0}{\sqrt{6}}} \rho ^2 B_1'+B_0 \left(\rho ^3 \phi_0'
\left(2 e^{\frac{7 \phi_0}{\sqrt{6}}} \phi_1'-3 \left(b_2+2 b_1
e^{\frac{7}{2} \sqrt{\frac{3}{2}} \phi_0}\right)
\phi_0'\right)\right.\right.\cr & \left.\left.+18 B_0' \left(2-7
\sqrt{6} \rho \phi_0'+e^{\frac{7}{2} \sqrt{\frac{3}{2}} \phi_0}
\left(4+7 \sqrt{6} \rho \phi_0'\right)\right)+6 B_0 \left(14
\sqrt{6} \left(-1+e^{\frac{7}{2} \sqrt{\frac{3}{2}} \phi_0}\right)
\phi_0'\right.\right.\right. \cr &+ \left.\left.\left. 49
\left(2+e^{\frac{7}{2} \sqrt{\frac{3}{2}} \phi_0}\right) \rho
\phi_0^{\prime 2}+e^{\frac{7}{2} \sqrt{\frac{3}{2}} \phi_0}
\left(c_1+d_1 e^{w_1 \phi_0}+d_2 e^{w_2\phi_0}+d_3
e^{w_3\phi_0}\right) \rho ^3 \phi_0^{\prime 4}\right.\right.\right.
\cr &+\left.\left.\left. 14 \sqrt{6} \left(-1+e^{\frac{7}{2}
\sqrt{\frac{3}{2}} \phi_0}\right) \rho
\phi_0''\right)\right)\right)=0,}}
\eqn\hteq{\eqalign{&2 e^{\frac{7 \phi_0}{\sqrt{6}}} \rho ^2\phi_1
\left(72 e^{\frac{\phi_0}{\sqrt{6}}} \rho +\sqrt{6}
\left(\phi_0'\left(6 \rho  B_0'+B_0 \left(18+\rho ^2 \phi_0^{\prime
2}\right)\right)+6 \rho B_0 \phi_0''\right)\right)\cr &+3 \left(\rho
B_0^2 \phi_0' \left(18 \left(32+3 e^{\frac{7}{2} \sqrt{\frac{3}{2}}
\phi_0} \left(5+8 c_1+8 d_1 e^{w_1\phi_0}+8 d_2 e^{w_2 \phi_0}+8 d_3
e^{w_3\phi_0}\right)\right) \rho \phi_0 ^{\prime 2}\right.\right.
\cr &+\left.\left.\left(56 \sqrt{6}+e^{\frac{7}{2}
\sqrt{\frac{3}{2}} \phi_0} \left(-56 \sqrt{6}+63 \sqrt{6} c_1+9
\left(d_1 e^{w_1\phi_0} \left(7 \sqrt{6}+12
w_2\right)\right.\right.\right.\right.\right. \cr
&+\left.\left.\left.\left.\left. d_2 e^{w_2\phi_0} \left(7
\sqrt{6}+12 w_2\right)+d_3 e^{w_3\phi_0} \left(7 \sqrt{6}+12
w_3\right)\right)\right)\right) \rho ^2 \phi_0^{\prime
3}\right.\right. \cr &+\left.\left. 48 e^{\frac{7}{2}
\sqrt{\frac{3}{2}} \phi_0} \left(c_1+d_1 e^{w_1\phi_0}+d_2
e^{w_2\phi_0}+d_3 e^{w_3\phi_0}\right) \rho ^3 \phi_0^{\prime 4}-84
\sqrt{6} \left(-1+e^{\frac{7}{2} \sqrt{\frac{3}{2}} \phi_0}\right)
\rho \phi_0''\right.\right. \cr &-\left.\left. 24 \phi_0'\left(7
\sqrt{6} \left(-1+e^{\frac{7}{2} \sqrt{\frac{3}{2}}
\phi_0}\right)-18 e^{\frac{7}{2} \sqrt{\frac{3}{2}} \phi_0}
\left(c_1+d_1 e^{w_1\phi_0}+d_2 e^{w_2\phi_0}+d_3
e^{w_3\phi_0}\right) \rho ^2 \phi_0''\right)\right)\right. \cr
&+\left. \rho \left(21 \sqrt{6} a_2 \rho ^2+36 B_0' \left(7 \sqrt{6}
B_0'-b_2 \rho ^2 \phi_0'\right)-3 e^{\frac{7}{2} \sqrt{\frac{3}{2}}
\phi_0} \left(7 \sqrt{6} a_1 \rho ^2+84 \sqrt{6} B_0^{\prime 2}+24
b_1 \rho ^2 B_0' \phi_0'\right)\right.\right. \cr &+\left.\left. 2
e^{\frac{7 \phi_0}{\sqrt{6}}} \rho \left(6 \rho B_0'\phi_1'+\phi_0'
\left(6 \rho B_1'+B_1\left(18+\rho ^2 \phi_0^{\prime
2}\right)\right)+6 \rho B_1\phi_0''\right)\right)\right. \cr
&-\left. 3 B_0 \left(2 B_0' \left(84 \sqrt{6}
\left(-1+e^{\frac{7}{2} \sqrt{\frac{3}{2}} \phi_0}\right)-\rho ^2
\phi_0^{\prime 2} \left(-35 \sqrt{6} \left(-1+e^{\frac{7}{2}
\sqrt{\frac{3}{2}} \phi_0}\right)\right.\right.\right.\right. \cr
&+\left.\left.\left.\left. 48 e^{\frac{7}{2} \sqrt{\frac{3}{2}}
\phi_0} \left(c_1+d_1 e^{w_1\phi_0}+d_2 e^{w_2\phi_0}+d_3
e^{w_3\phi_0}\right) \rho \phi_0'\right)\right)+\rho \left(-84
\sqrt{6} B_0''\right.\right.\right. \cr &- \left.\left.\left. 2
e^{\frac{7 \phi_0}{\sqrt{6}}} \rho \left(\phi_1' \left(6+\rho ^2
\phi_0^{\prime 2}\right)+2 \rho \phi_1''\right)+b_2 \rho \left(36
\phi_0'-7 \sqrt{6} \rho \phi_0^{\prime 2}+4 \rho ^2 \phi_0^{\prime
3}+12 \rho \phi_0''\right)\right.\right.\right.\cr
&+\left.\left.\left. e^{\frac{7}{2} \sqrt{\frac{3}{2}}\phi_0}
\left(84 \sqrt{6} B_0''+b_1 \rho \left(72 \phi_0'+7 \sqrt{6} \rho
\phi_0^{\prime 2}+8 \rho ^2 \phi_0^{\prime 3}+24 \rho
\phi_0''\right)\right)\right)\right)\right)=0,}}
\eqn\doeq{\eqalign{-\delta_1'&=\frac{1}{3} \rho  \phi_1'
\phi_0'+\frac{1}{6 \rho }e^{-\frac{7 \phi_0}{\sqrt{6}}} \left(-3 b_2
\rho ^2 \phi_0^{\prime 2}-6 b_1 e^{\frac{7}{2} \sqrt{\frac{3}{2}}
\phi_0} \rho ^2 \phi_0^{\prime 2}+288 B_0 \phi_0^{\prime 2}+135
e^{\frac{7}{2} \sqrt{\frac{3}{2}} \phi_0}B_0 \phi_0^{\prime
2}\right.\cr &-\left.21 \sqrt{6} \left(-1+e^{\frac{7}{2}
\sqrt{\frac{3}{2}} \phi_0}\right) \rho B_0 \phi_0^{\prime 3}+12
e^{\frac{7}{2} \sqrt{\frac{3}{2}} \phi_0} \left(c_1+d_1 e^{w_1
\phi_0}+d_2 e^{w_2 \phi_0}+d_3 e^{w_3 \phi_0}\right) \rho ^2 B_0
\phi_0^{\prime 4}\right.\cr &-\left.42 \sqrt{6} B_0 \phi_0''+42
\sqrt{6} e^{\frac{7}{2} \sqrt{\frac{3}{2}} \phi_0} B_0
\phi_0''\right).}}

The two-derivative solution can explicitly be written as
\eqn\backBd{B_0(\rho)=\frac{(3(r(\rho))^2+1)^2h(r(\rho))}{9(r(\rho))^2}\,,\;\phi_0(\rho)=\sqrt{\frac{2}{3}}\log\left(
1+\frac{1}{(r(\rho))^2}\right)\,,\;
\delta(\rho)=\log\frac{3(r(\rho))^2+1}{3\rho r(\rho)},}
where
\eqn\rrho{r(\rho)=\frac{-2\cdot 3^{1/3} +2^{1/3} \left(9 \rho
^3+\sqrt{12 +81 \rho ^6}\right)^{2/3}}{6^{2/3} \left(9 \rho
^3+\sqrt{12 +81 \rho ^6}\right)^{1/3}},}
is the solution to the equation
\eqn\rrhot{e^{2a(r(\rho))}=\rho^2\,.}

Expanding near the boundary, $\rho\gg 1$, we find
\eqn\nbphio{\phi_1=\frac{1}{\rho^2}(C_{f1}+C_{f2}\log\rho
+C_{fp}\log^2\rho)+\ldots,}
\eqn\nbbo{B_1=36\rho^2+\frac{C_{b2}}{\rho^2}+\frac{C_{b3}\log\rho}{\rho^2}+\frac{C_{bp}\log^2\rho}{\rho^2}\,+\ldots,}
where the expression \gbel\ plays a key role. The constants $C_{fp}$ and
$C_{bp}$ are independent of $r_H$ and $Q$. By choosing
\eqn\ABaobo{a_1=a_2=-120\,,\quad b_1=b_2=4,}
the constants $C_{fp}$ and $C_{bp}$ are zero. We have
verified this by fitting the numerical solution to the asymptotic
formulae \nbphio, \nbbo\ and making sure that the coefficients $C_{f1}$, $C_{f2}$, $C_{b2}$, $C_{b3}$ are stable
towards the change of the UV cut-off (the position of the boundary).

After we have solved for $B_1$ and $\phi_1$, we can use this solution to
integrate \doeq\ to find $\delta_1$.  We fix the constant of
integration in such a way that $g_{tt}/\rho^2=Be^{-2\delta}/\rho^2$
asymptotes to one near the boundary, so that the speed of light is equal
to one in the dual field theory. Using \nbphio\ and \nbbo\ we
obtain for $\rho\gg 1$
\eqn\deltanb{\delta_1=-\frac{-4+2 \left(-2169+12 b_1+6 b_2-2
\sqrt{6} C_{f1}\right) \beta }{36 \rho ^4}+18\beta +\ldots.}
Therefore the near-boundary expansion of $g_{tt}=Be^{-2\delta}$ is
given by
\eqn\gttnb{\frac{g_{tt}}{\rho^2}=\left(1+\frac{-1-2 \theta^2-\theta^4}{\rho
^4}\right)+\frac{\left(-1917+12 b_1+6 b_2+9 C_{b2}-2 \sqrt{6}
C_{f1}+648 \theta^2+324 \theta^4\right) \beta }{9 \rho ^4}\,.}

We want to extract $\delta g_{tt}^{(4)}$, the coefficient of the
sub-leading $1/\rho^4$ term at $\OO\left(\beta\right)$, as this will
allow us to determine the chemical potential
$A_t\left(\infty\right)$ from equation \ePrel.\foot{In equation
\ePrel\ we have
\eqn\encharge{r^5h'=-4\left(-(1+k^2)^2+\frac{\left(-1917+12 b_1+6
b_2+9 C_{b2}-2 \sqrt{6} C_{f1}+648 \theta^2+324 \theta^4\right)
\beta }{9 }\right)\,.}
}
At $\OO\left(\beta^0\right)$, this term, which
controls the energy density of the dual field theory, is
proportional to $Q^4$ at low $T$. It follows from \gttnb\ that, to
find the $\OO(\beta)$ correction to it, we need to find the
coefficients $C_{f1}$ and $C_{b2}$ in the near-boundary expansions
\nbphio\ and \nbbo\ of the dilaton and the metric.

Our numerical procedure is as follows. We solve equations \hoeq\ and \hteq\ for $\phi_1$ and
$B_1$, with the horizon boundary conditions given by
\eqn\rhco{B_1(\rho_H)=0\,,\quad \phi_1(\rho_H)=c\,,}
\eqn\rhct{\eqalign{\phi_1'(\rho _H)&= \frac{1}{12 \rho_H^2
B_0'(\rho_H)}e^{-\frac{7 \phi_0(\rho_H)}{\sqrt{6}}} \left(-21
\sqrt{6} a_2 \rho_H^2+21 \sqrt{6} a_1 e^{\frac{7}{2}
\sqrt{\frac{3}{2}} \phi_0(\rho_H)} \rho_H^2\right. \cr &-\left. 48
e^{4 \sqrt{\frac{2}{3}} \phi_0(\rho_H)} \rho_H^2 c-252 \sqrt{6}
B_0'(\rho_H)^{2}+252 \sqrt{6} e^{\frac{7}{2}
\sqrt{\frac{3}{2}} \phi_0(\rho_H)} B_0'(\rho_H)^{2}\right. \cr
&+\left. 6 a_2 \rho_H^3 \phi_0'(\rho_H)+12 a_1 e^{\frac{7}{2}
\sqrt{\frac{3}{2}} \phi_0(\rho_H)} \rho_H^3 \phi_0'(\rho_H)-16
\sqrt{6} e^{4 \sqrt{\frac{2}{3}} \phi_0(\rho_H)} \rho_H^3 c
\phi_0'(\rho_H)\right. \cr &+\left. 36 b_2 \rho_H^2
B_0'(\rho_H)\phi_0'(\rho_H)+72 b_1 e^{\frac{7}{2} \sqrt{\frac{3}{2}}
\phi_0(\rho_H)} \rho_H^2 B_0'(\rho_H) \phi_0'(\rho_H)\right).}\,}
We then fix numerically the value of $c$ such that $C_{f2}=C_{b3}=0$ (we
have verified numerically that both of these coefficients go to zero at
the same time for the particular value of $c$). This ensures that $\phi_1$ and
$B_1$ behave, near the boundary, in the same way as $\phi_0$ and $B_0$
respectively. For this value of $c$, we then extract numerically the coefficients
$C_{f1}$ and $C_{b2}$ in the near-boundary expansions.

From \gttnb\ we find that the $\OO(\beta)$ correction to the coefficient of $\rho^{-4}$ in $g_{tt}$ is
\eqn\coren{\frac{\delta g_{tt}^{(4)} }{Q^4} =-1917+12 b_1+6 b_2+9
C_{b2}-2 \sqrt{6} C_{f1}+648 \theta^2+324 \theta^4\,,}
and the $\OO\left(\beta\right)$ correction to the temperature is
\eqn\tempcor{\frac{\Delta T}{Q}=\tilde
T\left(\frac{B_1'(\rho_H)}{B_0'(\rho_H)}-\delta _1(\rho_H)\right),}
where $\tilde T=T/Q=\theta/\pi$ is the $Q$-normalized two-derivative
result for the temperature.

After fixing $w_1, w_2, w_3$, we find numerically that, for generic $c_1$, $d_{1,2,3}$, one finds (for $\theta\ll 1$),
\eqn\encorn{ \frac{\delta g_{tt}^{(4)}}{Q^4}=\frac{C_1}{\theta}+\OO(1),}
\eqn\Tcorn{ \frac{\Delta
T}{T}=\frac{C_2}{\theta^3}+\frac{C_3}{\theta^2}+\frac{C_4}{\theta}+\OO(1).}
By varying $c_1, d_1, d_2, d_3$, the four coefficients $C_{1,2,3,4}$ change. It is possible to make these coefficients numerically small, and we expect that by suitably fine tuning these coefficients, it is possible to produce a black hole solution with $C_{1,2,3,4}=0.$ However, this is a very expensive procedure to implement numerically and we have not been able to determine the required values of $c_1$ and $d_{1,2,3}$. Supposing that this solution exists, it would have $T\sim r_H$ and $\delta g_{tt}^{(4)}\sim Q^4$ to $\OO\left(\beta\right)$ in the small $T/\mu$ limit. Equation \ePrel\ then implies that $\mu\sim Q$ in this limit, as required.

\footatend\vfill\supereject\immediate\closeout\rfile\writestoppt
\baselineskip=14pt\centerline{{\bf References}}\bigskip{\frenchspacing%
\parindent=20pt\escapechar=` \input refs.tmp\vfill\eject}\nonfrenchspacing

\bye